\documentclass{iopjournal}

\usepackage{amssymb}
\usepackage{mathtools}
\usepackage{upgreek}
\usepackage{cite}

\begin{document}

\articletype{Paper} 

\title{Optimized matching conditions for self-guided laser wakefield accelerators}

\author{P. Valenta$^{1,*}$\orcid{0000-0003-1067-3762}, K. G. Miller$^2$\orcid{0000-0003-4826-9001}, B. K. Russell$^{3}$\orcid{0000-0003-0011-2419}, M. Lamač$^{1}$\orcid{0000-0001-9218-6073}, M. Jech$^{1, 4}$\orcid{0000-0002-6087-6771}, G. M. Grittani$^{1}$\orcid{0000-0003-4394-854X} and S. V. Bulanov$^{1}$\orcid{0000-0001-8305-0289}}

\affil{$^1$ELI Beamlines Facility, The Extreme Light Infrastructure ERIC, Dolní Břežany, Czech Republic}

\affil{$^2$Laboratory for Laser Energetics, University of Rochester, Rochester, NY, USA}

\affil{$^3$Department of Astrophysical Sciences, Princeton University, Princeton, NJ, USA}

\affil{$^4$Faculty of Information Technology, Czech Technical University in Prague, Prague, Czech Republic}

\affil{$^*$Author to whom any correspondence should be addressed.}

\email{petr.valenta@eli-beams.eu}


\begin{abstract}

We revisit the matching conditions for self-guided laser pulse propagation in plasma and refine their formulation to maximize the energy of electrons produced via laser wakefield acceleration. Bayesian optimization, combined with particle-in-cell simulations carried out in a quasi-three-dimensional geometry and a Lorentz-boosted frame, is employed. The optimization identifies the maximum electron energy that a self-guided laser wakefield accelerator, driven by a laser of a given energy, can produce, together with the corresponding acceleration distance. Our results further demonstrate that electrons with energies close to the maximum value can be obtained across a relatively wide range of input parameters and without the need for their precise tuning. This provides substantial flexibility for experimental implementation and significantly relaxes the operational constraints associated with self-guided laser wakefield accelerators.

\end{abstract}

\section{Introduction \label{sec:1}}

Laser wakefield acceleration (LWFA) has emerged as a rapidly advancing technique capable of achieving acceleration gradients several orders of magnitude higher than those attainable with conventional radio-frequency accelerators~\cite{tajima1979}. The underlying mechanism relies on the interaction between an intense laser pulse and a plasma, where the laser drives a large-amplitude plasma wave that can trap and accelerate electrons to relativistic energies \cite{esarey2009a}. Over recent decades, LWFA has attracted considerable attention owing to its potential to enable compact, high-energy electron accelerators \cite{aniculaesei2023, picksley2024} with broad applications, including advanced radiation sources~\cite{corde2013a, albert2014}, muon sources~\cite{zhang2025, ludwig2025, terzani2025a}, radiotherapy~\cite{svendsen2021a, horvath2023, guo2025}, and the exploration of strong-field quantum electrodynamics phenomena~\cite{gonoskov2022, russell2023}.

Despite remarkable progress, many aspects of the underlying physics of LWFA are still not fully understood, particularly those related to the control and optimization of the acceleration process. The main difficulty stems from the intrinsic complexity of LWFA, which simultaneously exhibits multi-physics, multi-scale, and multi-parametric behavior: it involves the coupled dynamics of electromagnetic fields, plasmas, and relativistic particles; physical processes that span scales from $ \mathrm{\upmu m} $ laser wavelengths to $ \mathrm{cm} $-scale acceleration lengths; and a strongly nonlinear dependence on numerous input parameters, where even small variations can lead to large, often non-intuitive changes in the resulting electron beam properties.

This complexity makes the two traditional pillars of physics—theory and experiment—particularly challenging to apply. While reduced analytical models and scaling laws offer valuable insights, they often fail to capture important features, especially in regimes relevant to next-generation applications. Experiments, on the other hand, are constrained by the capabilities of existing laser systems, the reproducibility of plasma targets, and the precision of diagnostics. Consequently, progress has often relied on trial-and-error approaches or narrow optimization around specific configurations, limiting the generality and scalability of the results obtained.

To address these challenges, research has increasingly turned to advanced computational and optimization techniques. Modern high-performance computing enables the modeling of laser–plasma interactions that capture the essential physical phenomena with high fidelity and reasonable computational cost. However, the number of simulations required to explore the high-dimensional and wide-ranging parameter space remains beyond current capabilities. To overcome this limitation, artificial intelligence and machine learning methods are playing an increasingly important role in efficiently navigating the parameter space, identifying optimal operating regimes, and reducing the total computational expense~\cite{dopp2023a, roussel2024}.

In the present work, we revisit the so-called matching conditions for self-guided laser pulse propagation in plasma~\cite{kostyukov2004, gordienko2005, lu2006, lu2007} and examine whether their formulation can be refined to enhance the maximum energy attainable by electrons accelerated through LWFA. At this stage, we restrict our analysis to the optimization of maximum electron energy alone, disregarding other beam parameters. This focus is motivated by specific applications in which exceeding a threshold energy is essential, such as muon production~\cite{zhang2025, ludwig2025, terzani2025a} and nuclear activation~\cite{nedorezov2021, kolenaty2022}. It is important to note, however, that the electron energy is expected to depend predominantly on the parameters of laser and the background plasma, whereas beam quality metrics (e.g., charge, energy spread, and divergence) are more strongly influenced by the specific electron injection mechanism. Consequently, the optimal conditions for maximizing electron energy identified here are expected to remain valid across different controllable injection schemes, while optimization of beam quality for a given injection mechanism can be performed subsequently.

To accomplish our goal, we employ Bayesian optimization (BO), a machine learning method particularly well suited for optimizing expensive-to-evaluate objective functions (i.e., functions to be minimized or maximized). In our study, the objective function—the maximum electron energy—is evaluated through a series of particle-in-cell (PIC) simulations using a quasi-three-dimensional (quasi-3D) geometry~\cite{lifschitz2009a, davidson2015} and a Lorentz-boosted frame~\cite{vay2007, fonseca2008, yu2016}. We note that BO has already been successfully applied to complex optimization problems in LWFA~\cite{shalloo2020, jalas2021, kirchen2021, irshad2024, valenta2025b}, and that both the quasi-3D geometry and the Lorentz-boosted-frame simulation techniques have been well established in previous studies~\cite{martins2010, yu2014, yu2016, massimo2025}.

We find that a self-guided LWFA driven by a $ 10~\mathrm{mJ} $ laser pulse can generate electrons with energies approaching $ 80~\mathrm{MeV} $ over an acceleration distance of less than $ 200~\mathrm{\upmu m} $. Importantly, electrons near this energy level can be produced through multiple combinations of input parameters and without requiring their precise tuning. This offers substantial flexibility for experimental realization and significantly relaxes the operational constraints of self-guided LWFA.

The paper is organized as follows. In Sec.~\ref{sec:2}, we introduce the key parameters of the Gaussian laser pulse. In Sec.~\ref{sec:3}, we discuss the fundamental physics of intense laser pulse propagation in plasma and revisit the matching conditions for self-guided propagation. In Sec.~\ref{sec:4}, we generalize the original formulation of the matching conditions and outline the assumptions used to minimize the number of parameters in the optimization process. Sections~\ref{sec:5} and~\ref{sec:6} describe the setups of the PIC simulations and the BO algorithm, respectively. In Sec.~\ref{sec:7}, we present the optimization results, and finally, in Sec.~\ref{sec:8}, we summarize our findings.

\section{Description of Gaussian laser pulse \label{sec:2}}

The spatiotemporal structure of a Gaussian laser pulse, commonly used as an idealized representation of real laser pulses, can be conveniently described in terms of its vector potential,
\begin{equation}\label{eq:gauss}
A = a \exp{(-i \varphi)} \hat{x}.
\end{equation}
Both the amplitude $ a(x, r, \xi) $ and the phase $ \varphi(x, r, t) $ depend on the longitudinal coordinate $ x $, the radial coordinate $ r $, the time $ t $, and the co-moving coordinate $ \xi = x - \beta c t $, which follows the laser pulse at its group velocity $ \beta c $, and $ \hat{x} $ denotes the unit vector along the laser propagation direction. For the sake of simplicity, only the $ x $ component of the vector potential is nonzero, corresponding to a radially polarized, cylindrically symmetric laser pulse. We consider the vector potential to be expressed in units of $ m_{\mathrm{e}} c / e $, where $ m_{\mathrm{e}} $ is the electron mass, $ c $ the speed of light in vacuum, and $ e $ the elementary charge.

The amplitude and phase can then be written as
\begin{equation}\label{eq:a}
a = a_0 \frac{x_{\mathrm{R}}}{\sqrt{x^2 + x_{\mathrm{R}}^2}} \exp{\left(-2\ln{2} \, \frac{\xi^2}{c^2 \beta^2 \tau_0^2} - \frac{r^2 x_{\mathrm{R}}^2}{w_0^2 \left( x^2 + x_{\mathrm{R}}^2 \right) }\right)},
\end{equation}
and
\begin{equation}\label{eq:psi}
\varphi = \omega_0 t - k_0 x \left(1 - \frac{r^2}{2 \left( x^2 + x_{\mathrm{R}}^2 \right)} \right) - \arctan{\left(\frac{x}{x_{\mathrm{R}}}\right)},
\end{equation}
respectively, where $ \omega_0 $ is the angular frequency of the pulse, $ k_0 = \omega_0 / c $ the wavenumber, $ \tau_0 $ the full width at half maximum (FWHM) duration of the intensity profile, $ a_0 $ the normalized amplitude, and $ w_0 $ the beam waist, i.e., the focal spot radius at which the intensity drops to $ 1 / \mathrm{e}^2 $ of its on-axis value. The Rayleigh length, $ x_{\mathrm{R}} = k_0 w_0^2 / 2 $, defines the distance from the focus to the point where the beam radius increases by a factor of $ \sqrt{2} $.

The total energy of a Gaussian laser pulse is given by
\begin{equation}\label{eq:ene_0}
\mathcal{E}_0 = \frac{\pi^{3/2}}{16 \sqrt{\ln{2}}} \, \overline{\mathcal{E}} \, a_0^2 \left(\frac{w_0}{\lambda_0}\right)^2 \tau_0 \omega_0,
\end{equation}
where $ \lambda_0 = 2 \pi c / \omega_0 $ is the laser wavelength,
\begin{equation}\label{eq:e_bar_p_bar}
\overline{\mathcal{E}} = \frac{\pi \overline{\mathcal{P}}}{\omega_0}, \quad \mathrm{and} \quad \overline{\mathcal{P}} = \frac{2 m_{\mathrm{e}} c^3}{r_{\mathrm{e}}}.
\end{equation}
Here, $ r_{\mathrm{e}} = e^2 / 4 \pi \epsilon_0 m_{\mathrm{e}} c^2 $ denotes the classical electron radius, with $ \epsilon_0 $ being the vacuum permittivity. The quantities in Eq.~(\ref{eq:e_bar_p_bar}) evaluate to $ \overline{\mathcal{E}} \approx 29~\mathrm{\upmu J} $ (for $ \lambda_0 = 1~\mathrm{\upmu m} $) and $ \overline{\mathcal{P}} \approx 17.4~\mathrm{GW} $. The corresponding laser power is then expressed as \cite{zeng2024}
\begin{equation}\label{eq:p_0}
\mathcal{P}_0 = 2 \sqrt{\frac{\ln{2}}{\pi}} \, \frac{\mathcal{E}_0}{\tau_0}.
\end{equation}

\section{Self-guiding and matching conditions \label{sec:3}}

For efficient LWFA, the laser pulse must preserve high intensity in plasma over distances significantly longer than those limited by diffraction. To counteract diffraction, guiding of the laser pulse within the plasma is essential. In general, this can be achieved either through external optical guiding or through self-guiding mechanisms. In the present work, we focus on the latter case. Self-guiding is experimentally attractive because it does not require precise control of external guiding structures, such as capillary-discharge waveguides~\cite{ehrlich1996, butler2002} or hydrodynamically formed plasma channels~\cite{durfee1993, miao2020}. On the other hand, it typically operates in a strongly nonlinear regime, which can lead to a reduction in the quality and stability of the accelerated electron beam compared with externally guided configurations \cite{esarey2009a, picksley2024}. These factors make self-guided LWFA a promising and relatively straightforward target for optimization.

The onset of self-guiding depends critically on the laser power relative to the threshold for relativistic self-focusing~\cite{sun1987},
\begin{equation}\label{eq:p_cr}
    \mathcal{P}_{\mathrm{cr}} = \overline{\mathcal{P}} \, \frac{n_{\mathrm{cr}}}{n_{\mathrm{e}}},
\end{equation}
where $ n_{\mathrm{e}} $ is the electron density and $ n_{\mathrm{cr}} = \pi / r_{\mathrm{e}} \lambda_0^2 $ is the critical plasma density. When the laser power is below the critical power $ \mathcal{P}_{\mathrm{cr}} $, diffraction dominates, causing the beam to expand and its amplitude to decrease. This limits the acceleration length and reduces the energy gain of trapped electrons. Conversely, when $ \mathcal{P}_0 \gg \mathcal{P}_{\mathrm{cr}} $, strong nonlinear effects (such as the laser filamentation instability~\cite{naumova2002, valenta2021a}, vortex generation~\cite{bulanov1996}, and soliton formation~\cite{bulanov1999, esirkepov2002}) can distort the beam profile and disrupt the acceleration process. Therefore, maintaining an appropriate ratio of the laser power to the critical power is essential for stable and efficient self-guided LWFA operation.

As an intense laser pulse propagates through an underdense plasma (i.e., $ n_{\mathrm{e}} < n_{\mathrm{cr}} $), its ponderomotive force expels electrons from the high-intensity region and leaves behind an ion cavity (often referred to as a ``bubble''~\cite{pukhov2002}). The charge separation creates a restoring force that pulls electrons back toward their initial positions, causing them to oscillate. The resulting structure—a sequence of ion cavities surrounded by thin, dense electron sheets—is often referred to as a ``wake'' and forms the basis of LWFA. The characteristic radius $ R $ of the ion cavity was originally estimated phenomenologically~\cite{kostyukov2004, gordienko2005, lu2006, lu2007} by balancing the laser ponderomotive force with the plasma restoring force,
\begin{equation}\label{eq:R}
    k_{\mathrm{p}} R \sim \sqrt{a_0},
\end{equation}
where $ k_{\mathrm{p}} = \omega_{\mathrm{p}} / c $ is the plasma wavenumber and $ \omega_{\mathrm{p}} = \sqrt{n_{\mathrm{e}} e^2 / m_{\mathrm{e}} \epsilon_0} $ is the plasma frequency. 

For efficient self-guiding, the laser spot size must be comparable to the size of the ion cavity, 
\begin{equation}\label{eq:w_0_matched}
    k_{\mathrm{p}} w_0 = 2 \sqrt{a_0},
\end{equation}
where the proportionality factor of $ 2 $, now widely accepted, has been ``inferred from simulations''~\cite{lu2006}. The power of a Gaussian laser pulse with a beam waist given by Eq.~(\ref{eq:w_0_matched}) is then $ \mathcal{P}_0 = \mathcal{P}_{\mathrm{cr}} \, a_0^3 / 8 $. Rearranging this expression for the normalized laser amplitude yields
\begin{equation}\label{eq:a_0_matched}
    a_0 = 2 \left( \frac{\mathcal{P}_0}{\mathcal{P}_{\mathrm{cr}}} \right)^{1/3}.
\end{equation}
Equations~(\ref{eq:w_0_matched}) and~(\ref{eq:a_0_matched}) are known as the matching conditions for self-guided laser propagation in plasma~\cite{kostyukov2004, gordienko2005, lu2006, lu2007}. They ensure that the plasma-induced focusing effect counteracts natural diffraction, enabling stable laser propagation in plasma over multiple Rayleigh lengths.

\section{Generalized matching conditions and parameterization \label{sec:4}}

In this work, we extend the standard formulation of the matching conditions by introducing a dimensionless proportionality parameter $ \kappa $ in place of the numerical factor $ 2 $ in Eq.~(\ref{eq:w_0_matched}), yielding
\begin{equation}\label{eq:w_0_matched_new}
    k_{\mathrm{p}} w_0 = \kappa \sqrt{a_0}.
\end{equation}
Accordingly, Eq.~(\ref{eq:a_0_matched}) can be rewritten as
\begin{equation}\label{eq:a_0_matched_new}
    a_0 = 2 \left( \frac{2}{\kappa} \right)^{2/3} \left( \frac{\mathcal{P}_0}{\mathcal{P}_{\mathrm{cr}}} \right)^{1/3}.
\end{equation}
The goal of this generalization is to determine whether there exists a value of $ \kappa $, different from its conventional value of 2, that maximizes the energy of electrons accelerated by LWFA within the first plasma period behind the driving laser pulse. To this end, we perform a systematic simulation study employing advanced numerical and optimization techniques that were not available at the time of the original studies.

To reduce the number of free parameters involved in the optimization process, we adopt the following simplifying assumptions: (i) LWFA is driven by a linearly polarized Gaussian laser pulse with a constant spectral phase, a specified central wavelength, and total energy; (ii) the plasma has a uniform density profile; (iii) the acceleration operates in the matched regime defined by Eqs.~(\ref{eq:w_0_matched_new}) and~(\ref{eq:a_0_matched_new}); and (iv) accelerated electrons are treated as test particles and are injected into the accelerating structure externally. These assumptions allow us to isolate the essential physical dependencies while minimizing computational complexity. It should be noted, however, that under these constraints the maximum achievable electron energy is not necessarily optimal. More complex configurations, such as LWFA driven by non-Gaussian~\cite{beaurepaire2015, oumbarekespinos2023} or frequency-chirped~\cite{kalmykov2012, kim2017a} laser pulses, tailored plasma profiles~\cite{bulanov1993, sprangle2001}, or operation in a mismatched regime~\cite{perevalov2020, poder2024}, can potentially yield higher energy gains, albeit at the cost of introducing additional degrees of freedom and complicating the optimization process. In future work, the methods developed in the present study may be extended to include these more general cases.

Under the assumptions (i)--(iv) outlined above, the LWFA process can be fully characterized by only three dimensionless parameters: $ \mathcal{P}_0 / \mathcal{P}_{\mathrm{cr}} $, $ \tau_0 \omega_{\mathrm{p}} $, and $ \kappa $. The corresponding laser and plasma quantities—namely, the pulse duration and amplitude, focal spot size, and plasma density—can then be expressed in the following normalized form:
\begin{equation}\label{eq:tau0_scaling}
    \tau_0 \omega_0 = 2 \sqrt{\pi \ln{2}} 
    \left( \frac{2 \sqrt{\pi \ln{2}}}{\tau_0 \omega_{\mathrm{p}}} \right)^{-2/3}
    \left( \frac{\mathcal{P}_0}{\mathcal{P}_{\mathrm{cr}}} \right)^{-1/3}
    \left( \frac{\mathcal{E}_0}{\overline{\mathcal{E}}} \right)^{1/3},
\end{equation}
\begin{equation}\label{eq:a0_scaling}
    a_0 = 2 \left( \frac{2}{\kappa} \right)^{2/3}
    \left( \frac{\mathcal{P}_0}{\mathcal{P}_{\mathrm{cr}}} \right)^{1/3},
\end{equation}
\begin{equation}\label{eq:w0_scaling}
    \frac{w_0}{\lambda_0} = \frac{\sqrt{2}}{\pi}
    \left( \frac{2}{\kappa} \right)^{-2/3}
    \left( \frac{2 \sqrt{\pi \ln{2}}}{\tau_0 \omega_{\mathrm{p}}} \right)^{1/3}
    \left( \frac{\mathcal{P}_0}{\mathcal{P}_{\mathrm{cr}}} \right)^{-1/6}
    \left( \frac{\mathcal{E}_0}{\overline{\mathcal{E}}} \right)^{1/3},
\end{equation}
\begin{equation}\label{eq:ne_scaling}
    \frac{n_{\mathrm{e}}}{n_{\mathrm{cr}}} =
    \left( \frac{2 \sqrt{\pi \ln{2}}}{\tau_0 \omega_{\mathrm{p}}} \right)^{-2/3}
    \left( \frac{\mathcal{P}_0}{\mathcal{P}_{\mathrm{cr}}} \right)^{2/3}
    \left( \frac{\mathcal{E}_0}{\overline{\mathcal{E}}} \right)^{-2/3}.
\end{equation}

In this study, we select a laser pulse energy of $ 10~\mathrm{mJ} $ and a wavelength of $ 1~\mathrm{\upmu m} $. Although the chosen energy is relatively low for typical LWFA experiments, it ensures that the corresponding acceleration length remains moderate (less than $ 200~\mathrm{\upmu m} $ in all simulated cases), thereby keeping the computational cost of the PIC simulations used for BO tractable. Furthermore, the selected laser parameters closely match the performance of state-of-the-art ultrafast fiber laser technology \cite{stark2021}, making the present results directly applicable to experimentally realizable setups. On the other hand, low-energy pulses in the context of LWFA require ultrashort durations, for which carrier–envelope phase effects can influence the wakefield dynamics and electron acceleration~\cite{nerush2009, valenta2020, salehi2021a, huijts2022, lazzarini2024}. The impact of these effects, as well as the corresponding scalability of the optimization to high-energy, long-pulse regimes, will be examined in future work.

\section{Particle-in-cell simulations \label{sec:5}}

The parameters of the PIC simulations are defined as follows. The driving laser pulse has an energy of $ 10~\mathrm{mJ} $ and a wavelength of $ 1~\mathrm{\upmu m} $. It has Gaussian spatial and temporal profiles [Eq.~(\ref{eq:a})], characterized by the FWHM duration $ \tau_0 $, normalized amplitude $ a_0 $, and beam waist $ w_0 $. The laser propagates along the $ x $ axis and is linearly polarized along the $ y $ axis. It travels through a fully ionized plasma slab with a uniform electron density $ n_{\mathrm{e}} $. To suppress plasma wave breaking and the associated spurious electron injection at the plasma–vacuum interface, a smooth 10-$\mathrm{\upmu m}$-long density ramp is applied to the front side of the slab. The focal spot of the laser pulse is located at the end of this density ramp. The values of $ \tau_0 $, $ a_0 $, $ w_0 $, and $ n_{\mathrm{e}} $ are iteratively determined according to Eqs.~(\ref{eq:tau0_scaling})--(\ref{eq:ne_scaling}), respectively, to find the optimal combination that maximizes the energy of the accelerated electrons, as described in more detail in Sec.~\ref{sec:6}.

To isolate the acceleration dynamics from the specifics of various injection mechanisms, an external beam of test electrons is introduced. The test electrons move self-consistently according to their charge and mass but do not contribute any current to the plasma, effectively acting as passive tracers. This approach is valid for regimes well below the beam-loading threshold~\cite{wilks1987, katsouleas1987}, where the influence of the accelerated electrons on the wakefield structure is negligible. The initial velocity of the test electrons is set to $ v_0 = c \sqrt{1 - \omega_{\mathrm{p}}^2 / \omega_0^2} $, approximately matching the phase velocity of the plasma wave to ensure efficient electron trapping. The beam is initialized as a cylinder, with length spanning the entire simulation domain and radius set to $ 2 w_0 $, thereby providing full overlap with the generated wakefield structure. Consequently, a continuous range of electron injection phases within the wakefield is sampled simultaneously in a single simulation. It is important to emphasize that the results obtained in the present manuscript remain valid for any experimentally relevant electron injection mechanism, provided that the approximate injection location and the injected charge can be controlled to a reasonable extent; an example of such a mechanism is nanoparticle-assisted injection \cite{cho2018, spadova2025}, which was employed in our previous study on the electron energy optimization \cite{valenta2025b}.

The PIC simulations are performed using \textsc{Osiris}~\cite{fonseca2002}. The simulations employ cylindrical geometry with an azimuthal Fourier decomposition of the electromagnetic field components (also referred to as quasi-3D geometry)~\cite{lifschitz2009a, davidson2015}. This configuration is well suited for systems exhibiting near-cylindrical symmetry, such as those encountered in LWFA, where a small number of azimuthal modes is sufficient to capture the essential physics. In this scheme, macroparticles (i.e., computational particles representing ensembles of nearby real particles in phase space) are allowed to move in three-dimensional space, while the electromagnetic fields are computed on a series of two-dimensional lattices, significantly reducing the computational cost. In our case, two azimuthal modes are included: the first accounts for axisymmetric plasma fields, and the second represents the non-axisymmetric, linearly polarized laser field.

The simulations employ a moving-window technique, in which the computational domain moves at the speed of light. The window dimensions are $ 24~\mathrm{\upmu m} $ and $ 16~\mathrm{\upmu m} $ in the longitudinal and radial directions, respectively. The underlying simulation grid is uniform, with resolutions of $ 30 $ and $ 15 $ cells per $ \lambda_0 $ in the longitudinal and radial directions, respectively. Each simulation is evolved over a duration of $ 200~T_0 $, where $ T_0 = \lambda_0 / c $ is the laser period. This time interval is sufficient to capture the maximum energy of the accelerated electrons in all cases studied.

To further reduce computational requirements, the simulations are performed in a Lorentz-boosted frame~\cite{vay2007, fonseca2008, yu2016} with a boost factor of $ \gamma_b = 2.6 $, determined to be optimal for the chosen configuration. Simulating in the Lorentz-boosted frame decreases the number of required time-steps; the computation of simulations within this work is about twice as fast as the corresponding lab-frame simulations.

The background plasma is assumed to be cold and collisionless and is represented by electron and ion macroparticles with cubic shape functions. Initially, each grid cell contains $ 32 $ background-electron, $ 32 $ test-electron, and $ 8 $ ion macroparticles. The equations of motion for the macroparticles are integrated using the standard Boris algorithm~\cite{boris1971}, while the electromagnetic fields are advanced with a customized finite-difference time-domain Maxwell solver~\cite{li2017} suitable for Lorentz-boosted frame simulations. 

\section{Bayesian optimization \label{sec:6}}

To efficiently explore the parameter space while minimizing the number of required PIC simulations, the simulations are coupled with a BO algorithm. For this purpose, we use \textsc{Optimas}, a framework designed for scalable optimization on high-performance computing systems~\cite{hudson2022, ferranpousa2023}.

The BO algorithm consists of two principal components: a probabilistic surrogate model and an acquisition function. The surrogate model approximates the objective function based on the available data, providing both a prediction and an uncertainty estimate in unexplored regions. The acquisition function then determines the next trial (i.e., an evaluation of the objective function at a specific point in the parameter space) by balancing the trade-off between exploration (i.e., sampling in less explored regions) and exploitation (i.e., refining the search in the vicinity of the current optimum). As the optimization proceeds, the surrogate model is continuously updated with new simulation results, yielding increasingly accurate predictions in subsequent iterations. This adaptive approach enables BO to efficiently identify optima with a significantly reduced number of evaluations compared with grid or random searches.

In our setup, a Gaussian process (GP)~\cite{rasmussen2005} is employed as the surrogate model, following standard practice in BO. A GP is defined as a collection of random variables for which any finite subset have a jointly Gaussian distribution. Formally, a function $ f(\mathbf{x}) $ drawn from a GP is denoted as
\begin{equation}
    f \left( \mathbf{x} \right) \sim \mathcal{GP} \left( m(\mathbf{x}), k(\mathbf{x}, \mathbf{x}^{\prime}) \right),
\end{equation}
where $ m(\mathbf{x}) $ is the mean function and $ k(\mathbf{x}, \mathbf{x}^{\prime}) $ is the covariance function (or kernel). The mean function represents the expected value of the function at the input $ \mathbf{x} $,
\begin{equation}
    m \left( \mathbf{x} \right) = \mathbb{E} \left[ f \left( \mathbf{x} \right) \right].
\end{equation}
The kernel specifies the correlation between function values evaluated at inputs $ \mathbf{x} $ and $ \mathbf{x}^{\prime} $,
\begin{equation}
    k \left( \mathbf{x}, \mathbf{x}^{\prime} \right) = \mathbb{E} \left[ \left( f \left( \mathbf{x} \right) - m \left( \mathbf{x} \right) \right) \left( f \left( \mathbf{x}^{\prime} \right) - m \left( \mathbf{x}^{\prime} \right) \right) \right].
\end{equation}

The choice of kernel reflects prior assumptions regarding the smoothness and overall behavior of the objective function. In this work, we adopt the radial basis function (RBF) kernel~\cite{powell1977},
\begin{equation}
    k \left( \mathbf{x}, \mathbf{x}^{\prime} \right) = \exp{\left( -\frac{1}{2} \left( \mathbf{x} - \mathbf{x}^{\prime} \right)^{\top} \Theta^{-2} \left( \mathbf{x} - \mathbf{x}^{\prime} \right) \right)},
\end{equation}
where $\Theta$ denotes the length-scale hyperparameter governing the rate at which correlations decay with distance in the input space. The value of $\Theta$ is determined by maximizing the log marginal likelihood of the GP. The RBF kernel assumes the objective function to be infinitely differentiable—an appropriate choice given the expected smooth dependence of the output on the input parameters.

The acquisition function employed is the Monte Carlo–based batched upper confidence bound ($q$-UCB)~\cite{auer2002, wilson2017}. The $q$-UCB acquisition function includes a tunable hyperparameter $ \beta_{q\text{-}\mathrm{UCB}} $, which governs the balance between exploration and exploitation. We set $ \beta_{q\text{-}\mathrm{UCB}} = 10 $ and keep it constant throughout the optimization process, as this value has proven to provide a good balance in a series of preliminary tests (see Appendix~A for details). Furthermore, a batched acquisition strategy allows multiple trials to be selected and evaluated concurrently.

Simulations considered within the present work yield noise-free observations. When extending the proposed approach to situations in which the objective function exhibits stochastic behavior, as occurs in experimental settings (e.g., when employing colliding-pulse injection \cite{gong2023}, where counter-propagating laser pulses induce stochastic electron heating), the Bayesian optimization framework must be adapted to explicitly account for observation noise.

Three input parameters are varied during optimization within the following fixed bounds: 
\begin{equation}\label{eq:parameter_space}
    1 \le \mathcal{P}_0 / \mathcal{P}_{\mathrm{cr}} \le 5, \quad 1 \le \tau_0 \omega_{\mathrm{p}} \le 5, \quad \mathrm{and} \quad 1.5 \le \kappa \le 2.5.
\end{equation}
Based on our previous work~\cite{valenta2025b}, the optimum is expected to lie within these intervals; hence, the parameter limits are not dynamically adjusted during the optimization process. The single optimization objective is the maximum test-electron macroparticle energy achieved in the simulation. Additionally, the acceleration length—defined as the propagation distance in plasma at which the maximum energy is reached—is recorded for post-analysis, although it is not included in the optimization objective. We do not attempt to distinguish among the individual mechanisms that determine the acceleration length (e.g., electron dephasing, pump depletion), as these processes are typically closely interconnected.

\begin{figure}
\centering
\includegraphics[width=0.9\textwidth]{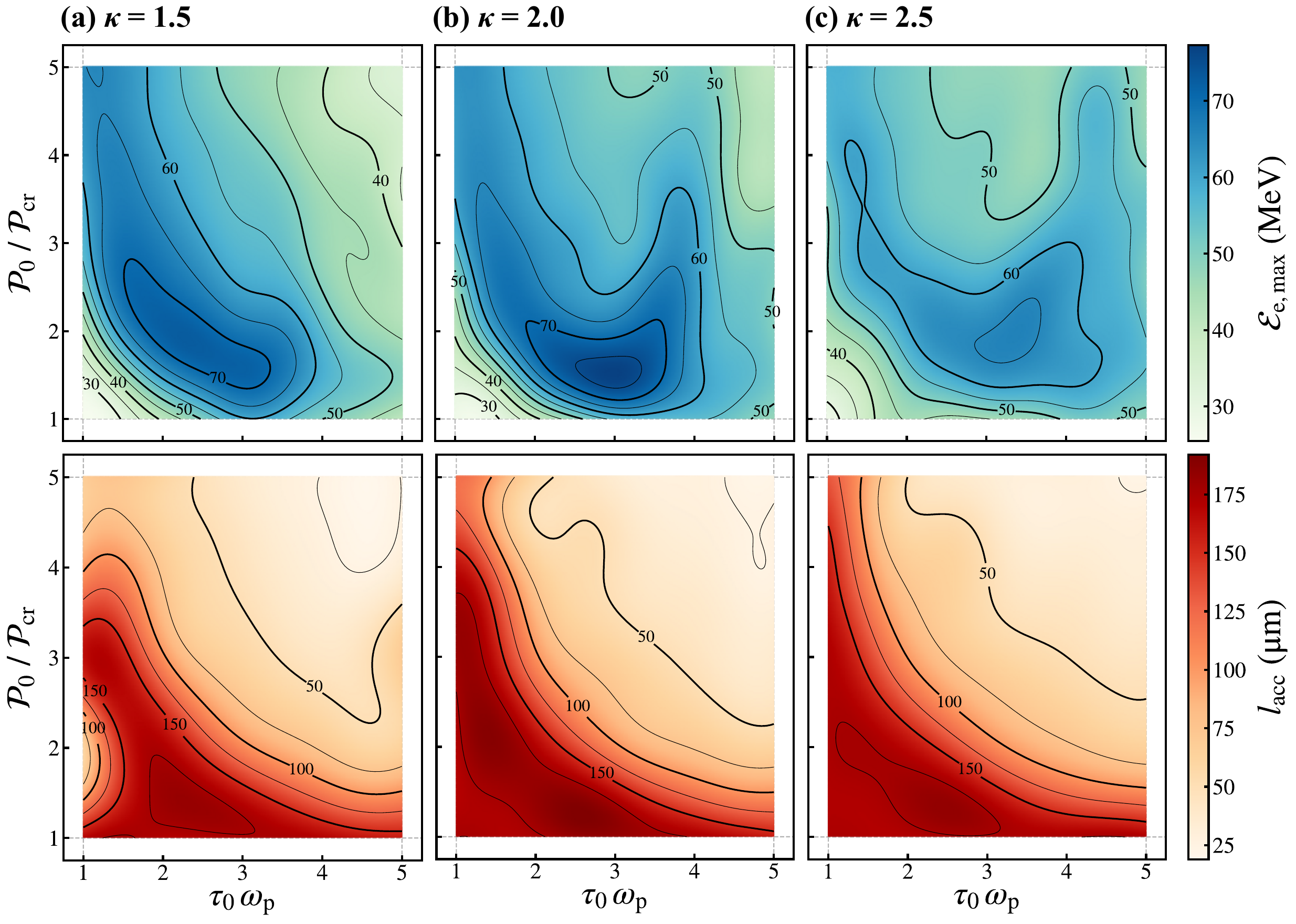}
\caption{Mean function of the GP model for the maximum electron energy, $ \mathcal{E}_{\mathrm{e, max}} $, and the corresponding acceleration distance, $ l_{\mathrm{acc}} $, based on the results of 500 PIC simulations for three different values of proportionality parameter $ \kappa $, (a) $ \kappa = 1.5 $, (b) $ \kappa = 2.0 $, and (c) $ \kappa = 2.5 $.}
\label{fig:1}
\end{figure}

The optimization workflow follows a distributed asynchronous scheme. A single manager process coordinates the generation and collection of trials, while eight worker processes perform independent trial evaluations in parallel. The asynchronous approach ensures that the optimization loop proceeds without waiting for all simulations to complete; this improves resource utilization given the nonuniform runtime of individual simulations. Communication between the manager and workers is handled via the Message Passing Interface (MPI)~\cite{clarke1994}. The manager runs on a dedicated compute node, and each simulation is executed on two nodes, with each node using 32 MPI processes and 4 threads per process.

To initialize the optimization, a set of eight quasi-random samples is generated using a Sobol sequence~\cite{sobol1967}. The optimization loop is terminated after a total budget of 500 trials. At this point, a sufficiently dense cluster of trials is obtained around the best candidate. Simulations over the final iterations around the estimated optimum show no significant variation, indicating that the search has converged.

\section{Optimization results \label{sec:7}}

Figure~\ref{fig:1} shows the mean function of the GP model for the maximum electron energy, $ \mathcal{E}_{\mathrm{e, max}} $, and the corresponding acceleration distance, $ l_{\mathrm{acc}} $, based on the results of 500 PIC simulations for $ \kappa = 1.5 $ [panel~(a)], $ \kappa = 2.0 $ [panel~(b)], and $ \kappa = 2.5 $ [panel~(c)]. The BO algorithm estimates the maximum electron energy to be $ \approx 77~\mathrm{MeV} $, reached over an acceleration distance of $ \approx 163~\mathrm{\upmu m} $. The optimal point in the parameter space is located at
\begin{equation}
    \mathcal{P}_0 / \mathcal{P}_{\mathrm{cr}} \approx 1.53, \quad \tau_0 \omega_{\mathrm{p}} \approx 2.96, \quad \mathrm{and} \quad \kappa \approx 2.06.
\end{equation}
This corresponds to the following set of initial laser and plasma parameters: $ a_0 \approx 2.3 $, $ \tau_0 \approx 9.5~\mathrm{fs} $ (yielding $ \mathcal{P}_0 \approx 1~\mathrm{TW} $), $ w_0 \approx 3~\mathrm{\upmu m} $, and $ n_{\mathrm{e}} \approx 3 \times 10^{19}~\mathrm{cm^{-3}} $. Across the explored parameter space, the maximum electron energy ranges from $ \approx 25 $ to $ 77~\mathrm{MeV} $, with the corresponding acceleration distance varying between $ \approx 19 $ and $ 192~\mathrm{\upmu m} $. For additional information on the uncertainty quantification of the GP models and their comparison with classical analytical models, see Appendices B and C, respectively.

\begin{figure}
\centering
\includegraphics[width=0.6\textwidth]{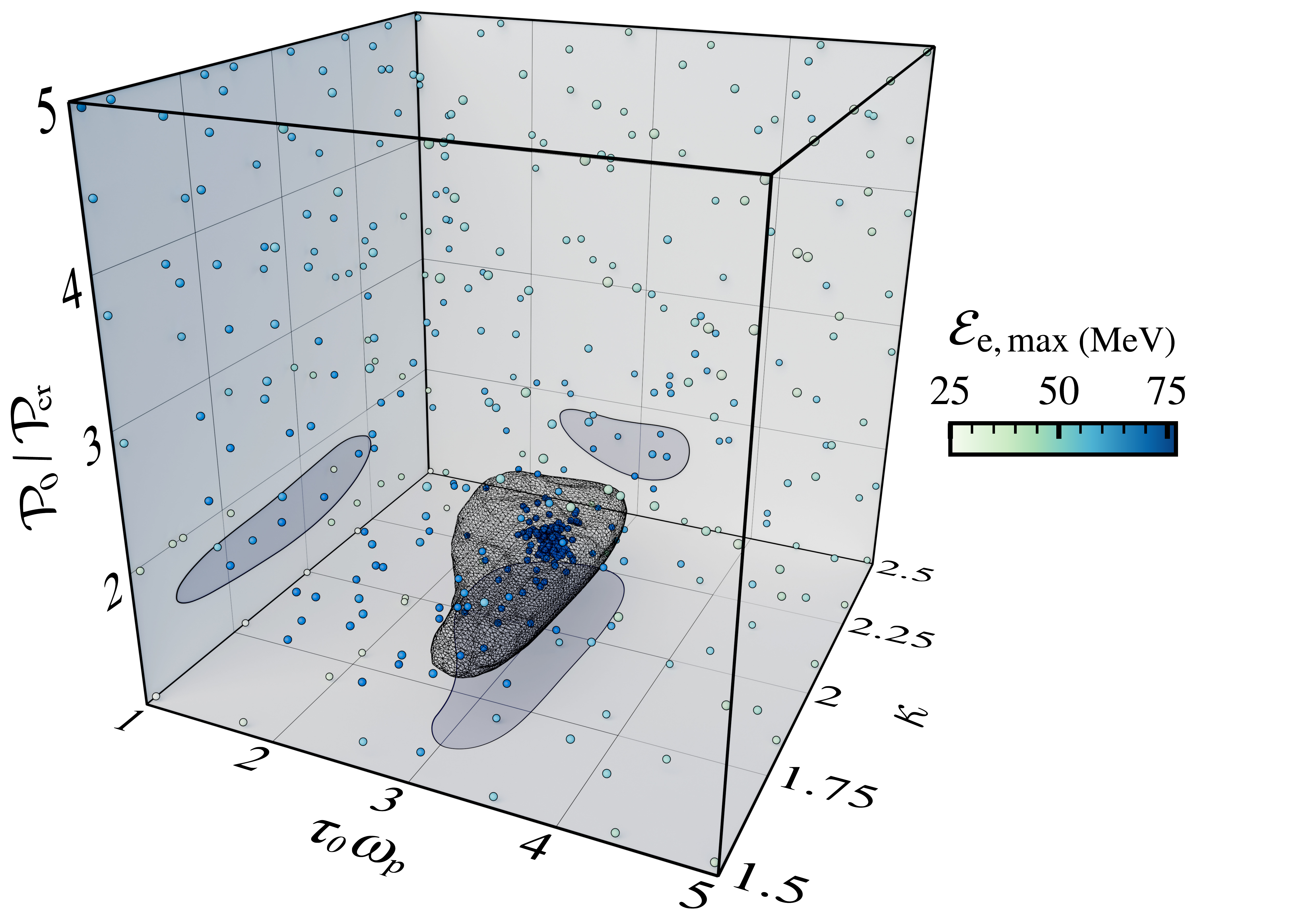}
\caption{Locations of the trials in the parameter space and the corresponding maximum electron energies, $ \mathcal{E}_{\mathrm{e,max}} $. The black mesh encloses the region where the upper $ 5\% $ of the maximum electron energy distribution (i.e., $ \mathcal{E}_{\mathrm{e,max}} > 73.5~\mathrm{MeV} $) is obtained. The blue areas show the projections of this region onto the respective coordinate planes.}
\label{fig:2}
\end{figure}

Figure~\ref{fig:2} presents the locations of the trials in the parameter space together with their corresponding $ \mathcal{E}_{\mathrm{e,max}} $. It also highlights the region where the upper $ 5\% $ of the maximum electron energy distribution (i.e., $ \mathcal{E}_{\mathrm{e,max}} > 73.5~\mathrm{MeV} $) is obtained. This region, which delineates the optimal operating regime of LWFA, spans a relatively broad portion of the parameter space bounded by
\begin{equation}\label{eq:optimal_region}
    1.3 < \mathcal{P}_0 / \mathcal{P}_{\mathrm{cr}} < 1.8, \quad 2.3 < \tau_0 \omega_{\mathrm{p}} < 3.5, \quad \mathrm{and} \quad 1.6 < \kappa < 2.3.
\end{equation}
This indicates that near-maximum electron energies can be produced across several different sets of initial parameters. For instance, in the present $ 10~\mathrm{mJ} $ case, electrons in the top $ 5\% $ of the energy distribution are obtained either for $ \tau_0 \approx 9.8~\mathrm{fs} $, $ w_0 \approx 2.4~\mathrm{\upmu m} $, and $ n_{\mathrm{e}} \approx 3.4 \times 10^{19}~\mathrm{cm^{-3}} $, or for $ \tau_0 \approx 8.1~\mathrm{fs} $, $ w_0 \approx 3.3~\mathrm{\upmu m} $, and $ n_{\mathrm{e}} \approx 2.9 \times 10^{19}~\mathrm{cm^{-3}} $. Furthermore, by fixing the pulse duration and plasma density near their optimal values (i.e., $ \tau_0 \approx 9.5~\mathrm{fs} $ and $ n_{\mathrm{e}} \approx 3 \times 10^{19}~\mathrm{cm^{-3}} $), variations of the focal spot from $ \approx 2.4 $ to $ 3.3~\mathrm{\upmu m} $ still yield near-maximum electron energies. These findings demonstrate that operation within the optimal LWFA regime can be (i) realized through multiple parameter combinations, providing substantial flexibility for experimental implementation, and (ii) maintained without the need for precisely tuned laser or plasma parameters, thereby significantly relaxing the operational constraints.

\begin{figure}
\centering
\includegraphics[width=1\textwidth]{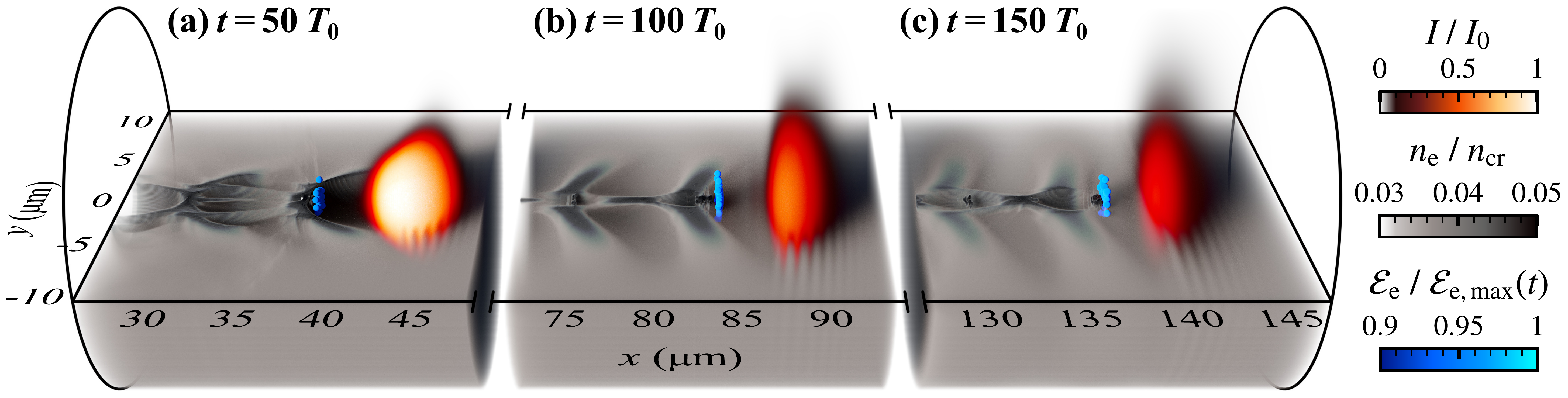}
\caption{Spatial distributions of the laser intensity, $ I $, electron density, $ n_{\mathrm{e}} $, and energy of accelerated electrons, $ \mathcal{E}_{\mathrm{e}} $, at three successive time instants, (a) $ t = 50 \, T_0 $, (b) $ t = 100 \, T_0 $, and (c) $ t = 150 \, T_0 $, obtained from the PIC simulation with input parameters closest to the optimum identified by BO. Only test-electron macroparticles with energies within the upper $ 10\% $ of the maximum electron energy at the given time step, $ \mathcal{E}_{\mathrm{e,max}}(t) $, are shown and $ I_0 = 2 \mathcal{P}_0 / \pi w_0^2 $. The plasma density is sliced along the $ x $–$ y $ plane to reveal the inner structure of the plasma wave.}
\label{fig:3}
\end{figure}

Figure~\ref{fig:3} illustrates the spatial distributions of the laser intensity, electron density, and the most energetic test-electron macroparticles at three successive time instants, obtained from the PIC simulation with input parameters closest to the optimum identified by BO. At $ t = 50~T_0 $ [panel~(a)], the diameter of the ion cavity within the first plasma wave period is comparable to the transverse size of the laser pulse, confirming stable self-guiding. At $ t = 100~T_0 $ [panel~(b)], the laser energy begins to deplete, and the electric field associated with self-injected electrons (not included in the optimization process) starts to interfere with the accelerating field. By $ t = 150~T_0 $ [panel~(c)], the laser energy is nearly depleted, and the acceleration process is about to cease. 

The simulation results indicate that achieving efficient self-guiding under realistic conditions requires accounting for the self-consistent evolution of the laser pulse and the plasma. As discussed in Ref.~\citenum{valenta2025b}, this implies that (i) the initial laser power should be set slightly above the critical power and (ii) the pulse duration be somewhat longer than the resonant duration for linear plasma wave excitation. We note that, for a Gaussian pulse with $ a_0 \ll 1 $, the resonant duration can be calculated analytically as $ \tau_r \omega_{\mathrm{p}} \approx 2 \sqrt{2 \ln{2}} \approx 2.35 $~\cite{leemans1996}.

In this work, we build upon our previous study~\cite{valenta2025b}, where BO was applied to the case of $ \kappa = 2 $, the electron injection was realized using the nanoparticle-assisted method~\cite{cho2018, spadova2025}, and all other laser and plasma parameters were kept identical. The adoption of advanced numerical techniques in the present analysis enabled a more thorough exploration of the parameter space, resulting in a maximum electron energy almost 15\% higher than reported in Ref.~\citenum{valenta2025b}.

\section{Conclusion \label{sec:8}}

We have revisited the matching conditions for self-guided laser pulse propagation in underdense plasma and extended their original formulation by introducing a dimensionless proportionality parameter that relates the laser beam waist to the radius of the ion cavity. The purpose of this modification was to identify the optimal value of this parameter for maximizing the energy of electrons accelerated via LWFA. To accomplish this, we employed BO in combination with PIC simulations carried out in a quasi-3D geometry and a Lorentz-boosted frame to reduce computational cost. 

The optimization revealed that a self-guided LWFA driven by a $ 10~\mathrm{mJ} $ laser pulse in a uniform plasma is, in principle, capable of producing electrons with energies approaching $ 80~\mathrm{MeV} $ over an acceleration distance below $ 200~\mathrm{\upmu m} $. The maximum energy was achieved for a proportionality parameter close to the widely accepted factor of $ 2 $ ($\pm0.4$). The underlying physics is straightforward: efficient self-guiding requires a balance between the plasma restoring force and the laser ponderomotive force. Furthermore, electrons with energies near the maximum value were obtained across a relatively broad range of input parameters and without the need for their precise tuning, i.e., the balance between these forces need not be exact because LWFA is a dynamic, self-adjusting process. For instance, when the pulse duration and plasma density are set to $ 9.5~\mathrm{fs} $ and $ 3 \times 10^{19}~\mathrm{cm^{-3}} $, respectively, varying the focal spot between $ 2.4 $ and $ 3.3~\mathrm{\upmu m} $ still yields near-maximum electron energies. This provides substantial flexibility for experimental implementation and significantly relaxes the operational constraints associated with self-guided LWFA. Although the optimization is demonstrated for a relatively low-energy driver pulse, the parameter space is defined in terms of dimensionless quantities, for which the location and extent of the optimal operating region are expected to depend only weakly on the laser energy due to the self-similar nature of the underlying physics. 

In summary, this work presents a systematic numerical investigation that integrates advanced simulation techniques with machine-learning–based optimization to validate and refine the matching conditions for self-guided LWFA. The results obtained provide a solid foundation for future theoretical and experimental efforts, including extensions toward higher-energy, collider-relevant parameter regimes.


\funding{This work was supported by the NSF and Czech Science Foundation (NSF-GACR collaborative Grant No.~2206059 and Czech Science Foundation Grant No.~22-42963L) and by the project ``e-INFRA CZ'' (ID:90254) from the Ministry of Education, Youth and Sports of the Czech Republic. This material is based upon work supported by the Office of Fusion Energy Sciences under Award Numbers DE-SC0021057, the Department of Energy (DOE) [National Nuclear Security Administration (NNSA)] University of Rochester ``National Inertial Confinement Fusion Program'' under Award Number DE-NA0004144. B.K.R. was supported by the US Department of Energy High-Energy-Density Laboratory Plasma Science program under Grant No.~DE-SC0020103.}


\data{The data that support the findings of this study are openly available at the following URL/DOI: \url{https://doi.org/10.5281/zenodo.18610970}. \cite{valenta2026_zenodo}}


\section*{Appendix A. Acquisition function hyperparameter tuning}

The hyperparameter $\beta_{q,\mathrm{UCB}}$ of the $q$-UCB acquisition function controls the trade-off between exploration and exploitation in the Bayesian optimization procedure, and there is no universally optimal choice for its value. Instead, the appropriate setting depends on factors such as the available evaluation budget and the specific optimization objective. For example, in scenarios with a very limited number of evaluations (e.g., costly experiments or real-time tuning), it is often preferable to emphasize exploitation in order to identify a satisfactory solution rapidly, even at the risk of missing the global optimum. Conversely, in cases where robustness or the global structure of the objective landscape is of interest, stronger exploration may be favored.

\begin{figure}
\centering
\includegraphics[width=0.8\textwidth]{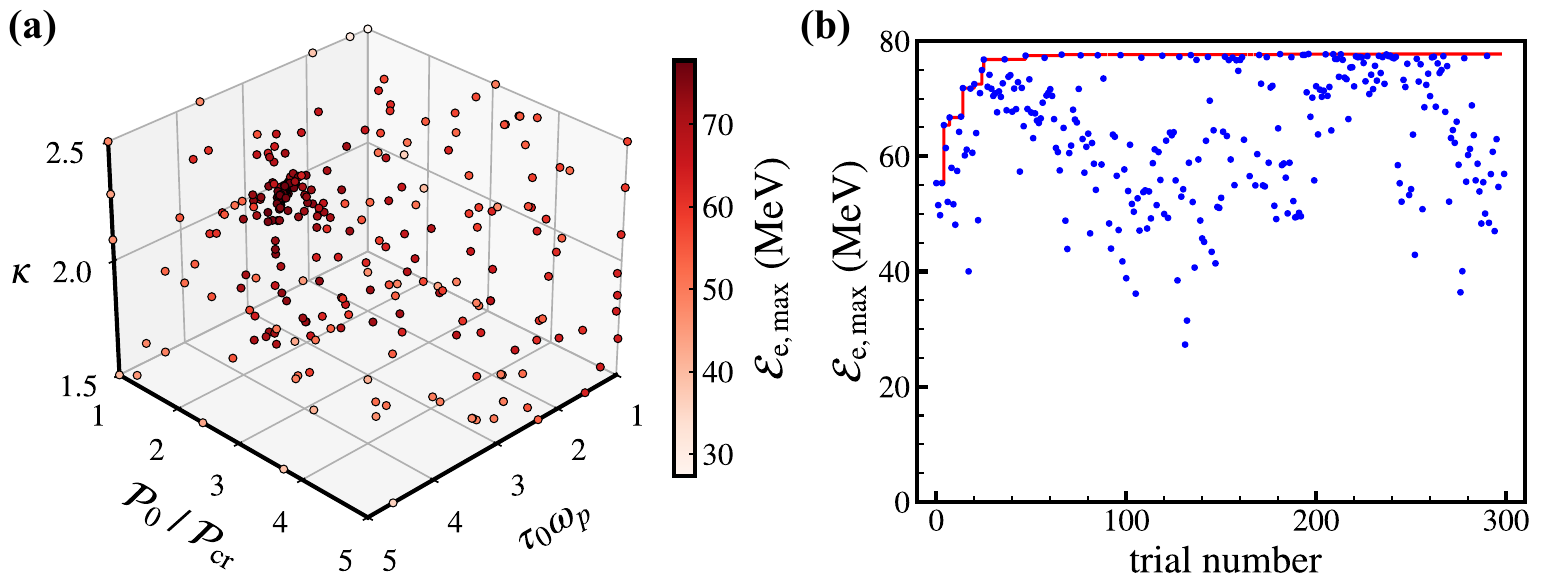}
\caption{(a) Locations of the trials in the parameter space and the corresponding maximum electron energies, $ \mathcal{E}_{\mathrm{e,max}} $, and (b) maximum electron energy obtained in each evaluation (blue) and the evolution of the cumulative maximum (red) for $\beta_{q,\mathrm{UCB}} = 1$.}
\label{fig:4}
\end{figure}

\begin{figure}
\centering
\includegraphics[width=0.8\textwidth]{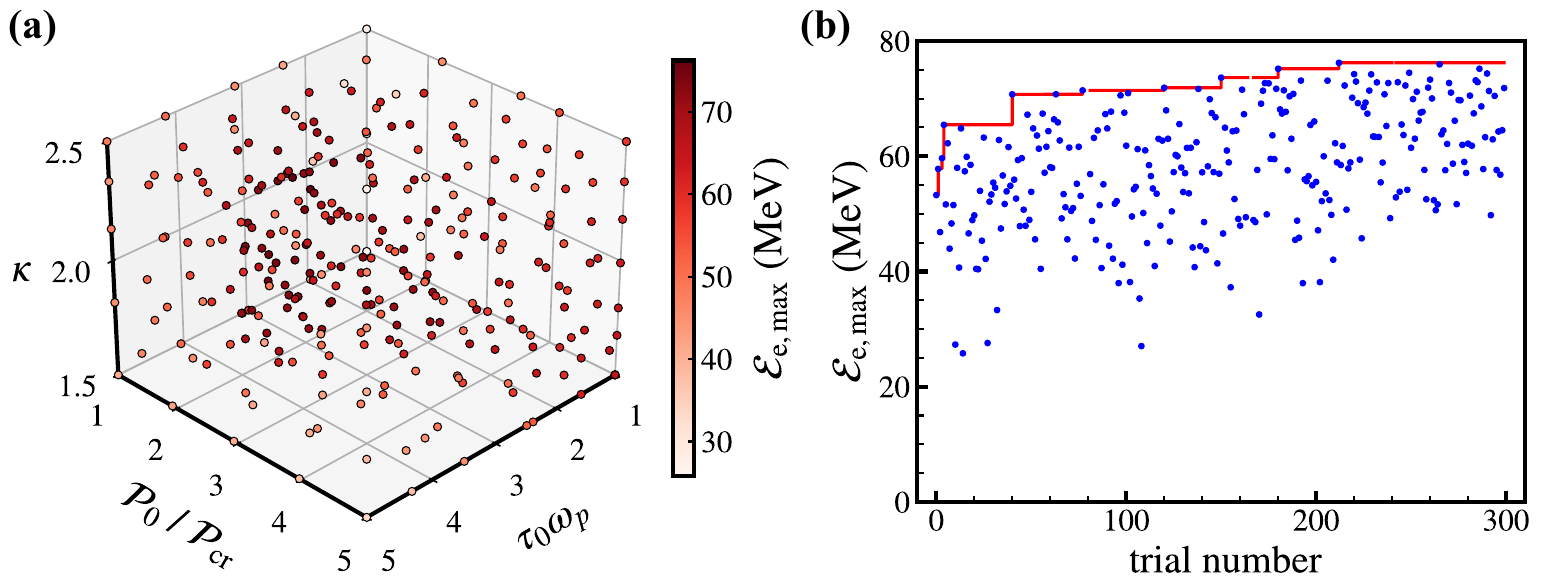}
\caption{(a) Locations of the trials in the parameter space and the corresponding maximum electron energies, $ \mathcal{E}_{\mathrm{e,max}} $, and (b) maximum electron energy obtained in each evaluation (blue) and the evolution of the cumulative maximum (red) for $\beta_{q,\mathrm{UCB}} = 100$.}
\label{fig:5}
\end{figure}

In the present study, we employ a relatively large evaluation budget of 500 PIC simulations, which allows for a balanced strategy combining both exploration and exploitation. Based on preliminary tests, we found $\beta_{q,\mathrm{UCB}} = 10$ to provide a suitable compromise under these conditions. To support this choice, we include additional results showing optimization runs with $\beta_{q,\mathrm{UCB}} = 1$ (Fig.~\ref{fig:4}) and $\beta_{q,\mathrm{UCB}} = 100$ (Fig.~\ref{fig:5}), using an evaluation budget of 300 PIC simulations. These results demonstrate that smaller values of $\beta_{q,\mathrm{UCB}}$ lead to rapid convergence toward local optima with limited exploration of the parameter space, whereas larger values overly emphasize exploration at the expense of convergence. The chosen value $\beta_{q,\mathrm{UCB}} = 10$ therefore represents a practical and well-balanced choice for the optimization problem considered here.

\section*{Appendix B. Uncertainty quantification of surrogate models}

\begin{figure}
\centering
\includegraphics[width=0.9\textwidth]{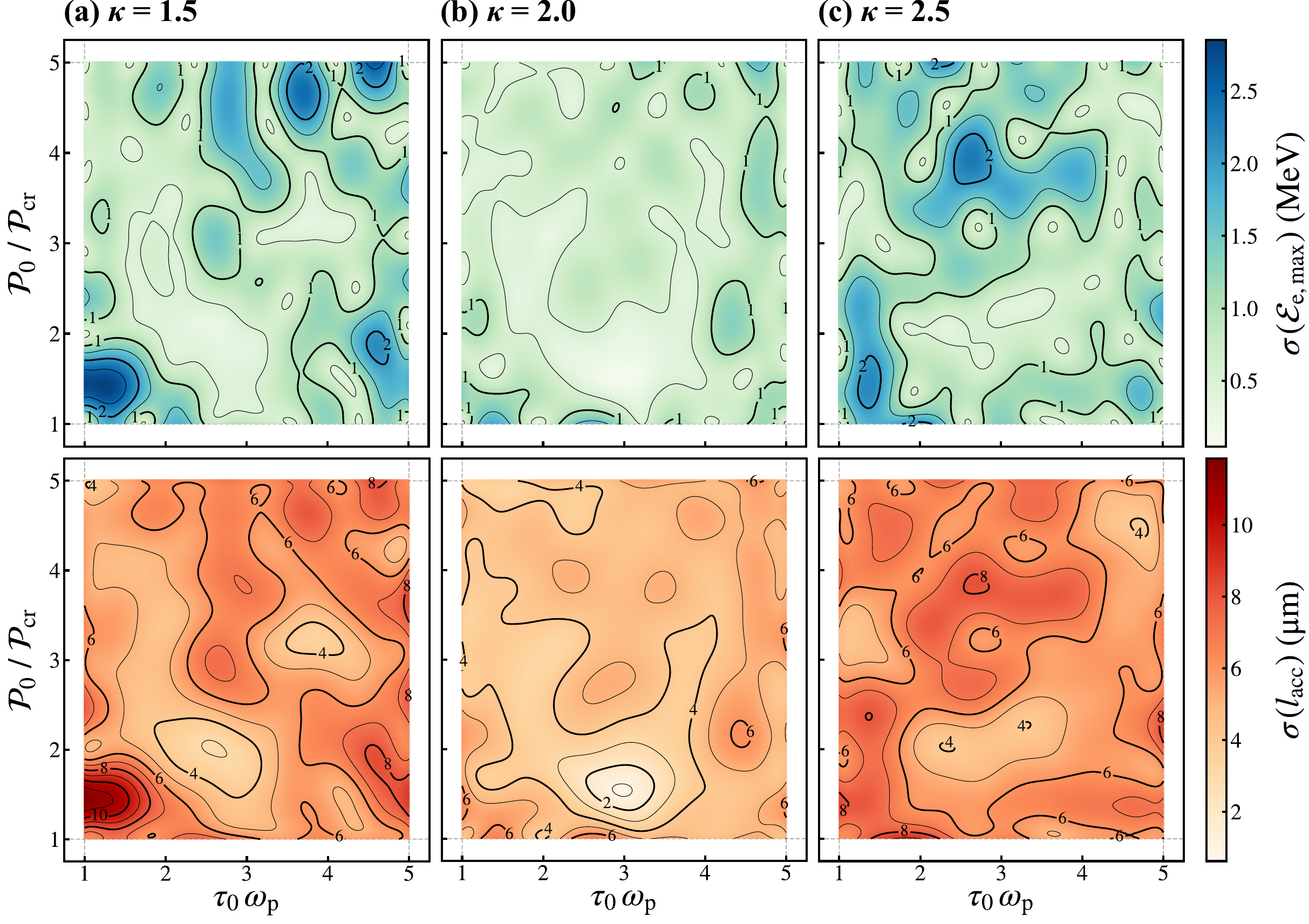}
\caption{Standard deviation of the GP model for the maximum electron energy, $ \sigma (\mathcal{E}_{\mathrm{e, max}}) $, and the corresponding acceleration distance, $ \sigma (l_{\mathrm{acc}}) $, based on the results of 500 PIC simulations for three different values of proportionality parameter $ \kappa $, (a) $ \kappa = 1.5 $, (b) $ \kappa = 2.0 $, and (c) $ \kappa = 2.5 $.}
\label{fig:6}
\end{figure}

GP models quantify predictive uncertainty through the posterior standard deviation, which provides a fundamental measure of confidence in the model predictions. A high standard deviation indicates large uncertainty, typically occurring in regions far from the training data, whereas a low standard deviation corresponds to high predictive confidence near sampled points.

To illustrate the spatial distribution of predictive uncertainty, Fig.~\ref{fig:6} shows the standard deviation of the GP model for the maximum electron energy, $ \sigma(\mathcal{E}_{\mathrm{e,max}}) $, and the corresponding acceleration distance, $ \sigma(l_{\mathrm{acc}}) $ for $ \kappa = 1.5 $ [panel~(a)], $ \kappa = 2.0 $ [panel~(b)], and $ \kappa = 2.5 $ [panel~(c)]. Within the parameter space defined by Eq.~(\ref{eq:parameter_space}), the maximum values of $ \sigma(\mathcal{E}_{\mathrm{e,max}}) $ and $ \sigma(l_{\mathrm{acc}}) $ are $ \approx 2.9\,\mathrm{MeV} $ and $ 11.9\,\mathrm{\upmu m} $, respectively. In the optimal LWFA operating regime defined by Eq.~(\ref{eq:optimal_region}), the standard deviation of the predicted maximum electron energy remains below \( 1\,\mathrm{MeV} \), indicating that the excellent performance predicted in this region is robust and supported by high model confidence.

\section*{Appendix C. Comparison with classical analytical models}

\begin{figure}
\centering
\includegraphics[width=0.75\textwidth]{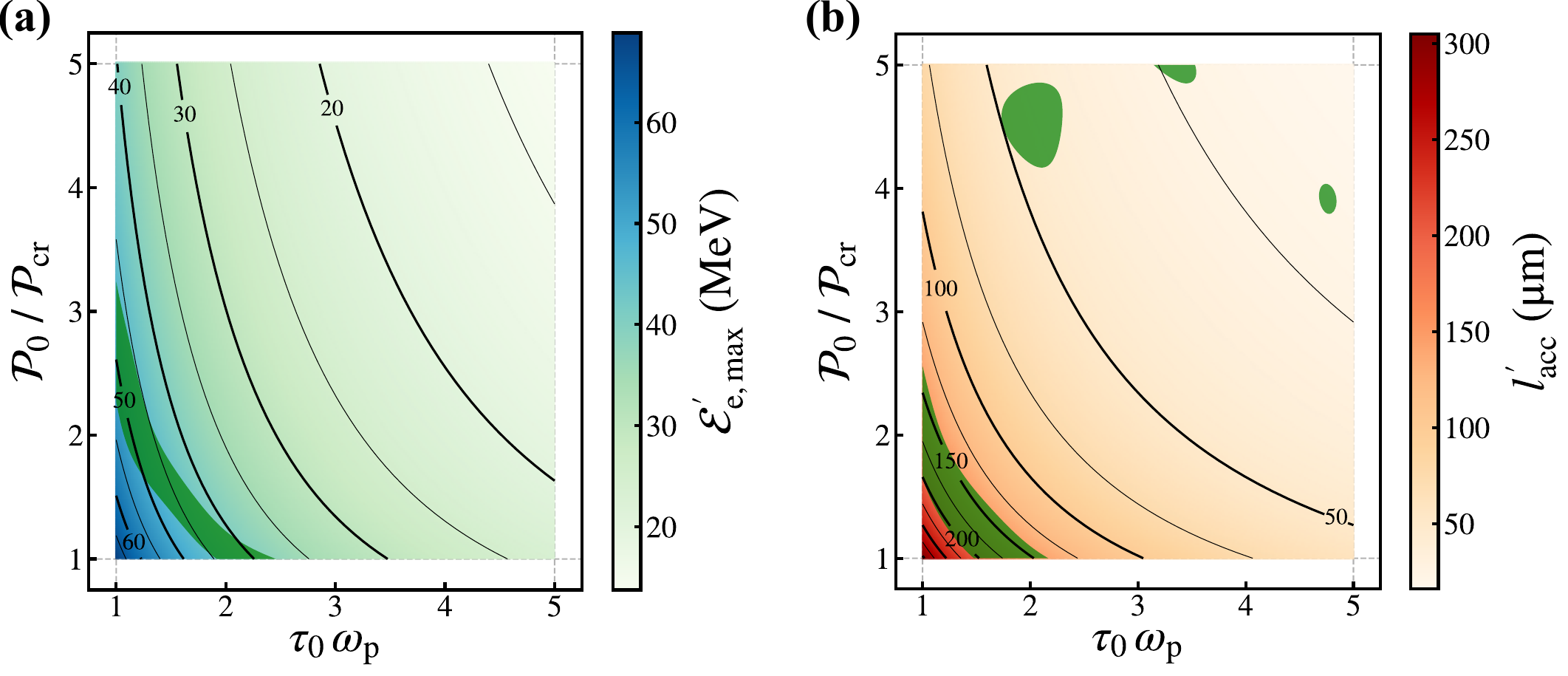}
\caption{Analytical model predictions for the maximum electron energy, \( \mathcal{E}_{\mathrm{e,max}}^{\prime} \), and the corresponding acceleration distance, \( l_{\mathrm{acc}}^{\prime} \), as presented in Ref.~\citenum{lu2007}. The green regions indicate parameter-space areas where the analytical model agrees with the GP surrogate model shown in panel~(b) of Fig.~\ref{fig:1} to within a 20\% relative error.}
\label{fig:7}
\end{figure}

To place the results obtained in this work within the context of current physical understanding, we compare the GP surrogate models for the maximum electron energy and the corresponding acceleration distance with the classical, well-established analytical expressions presented in Ref.~\citenum{lu2007}. In that work, the maximum electron energy and the corresponding acceleration distance are given by
\begin{equation}\label{eq:lu_ene}
    \frac{\mathcal{E}_{\mathrm{e,max}}^{\prime}}{m_{\mathrm{e}} c^2} \approx \frac{2}{3} \left( \frac{n_{\mathrm{e}}}{n_{\mathrm{cr}}} \right)^{-1} a_0
\end{equation}
and
\begin{equation}\label{eq:lu_lacc}
    k_0 l_{\mathrm{acc}}^{\prime} \approx \frac{4}{3} \left( \frac{n_{\mathrm{e}}}{n_{\mathrm{cr}}} \right)^{-3/2} \sqrt{a_0},
\end{equation}
respectively. Since Eqs.~(\ref{eq:lu_ene}) and (\ref{eq:lu_lacc}) are derived in the matched regime, we use Eqs.~(\ref{eq:a0_scaling}) and (\ref{eq:ne_scaling}) (with \( \kappa = 2 \)) to express them, without loss of generality, in terms of the dimensionless parameters \( \mathcal{P}_0 / \mathcal{P}_{\mathrm{cr}} \) and \( \tau_0 \omega_{\mathrm{p}} \):
\begin{equation}\label{eq:lu_ene_transform}
    \frac{\mathcal{E}_{\mathrm{e,max}}^{\prime}}{m_{\mathrm{e}} c^2} \approx \frac{4}{3}
    \left( \frac{2 \sqrt{\pi \ln{2}}}{\tau_0 \omega_{\mathrm{p}}} \right)^{2/3}
    \left( \frac{\mathcal{P}_0}{\mathcal{P}_{\mathrm{cr}}} \right)^{-1/3}
    \left( \frac{\mathcal{E}_0}{\overline{\mathcal{E}}} \right)^{2/3},
\end{equation}
\begin{equation}\label{eq:lu_lacc_transform}
    k_0 l_{\mathrm{acc}}^{\prime} \approx \frac{4 \sqrt{2}}{3}
    \left( \frac{2 \sqrt{\pi \ln{2}}}{\tau_0 \omega_{\mathrm{p}}} \right)
    \left( \frac{\mathcal{P}_0}{\mathcal{P}_{\mathrm{cr}}} \right)^{-5/6}
    \left( \frac{\mathcal{E}_0}{\overline{\mathcal{E}}} \right).
\end{equation}
Equations~(\ref{eq:lu_ene_transform}) and (\ref{eq:lu_lacc_transform}) are plotted in Fig.~\ref{fig:7}, together with the regions in which they agree with the GP model shown in panel~(b) of Fig.~\ref{fig:1} to within a 20\% relative error. It can be seen that Eqs.~(\ref{eq:lu_ene}) and (\ref{eq:lu_lacc}) are well applicable along narrow stripes in the parameter space, corresponding to values of \( \mathcal{P}_0 / \mathcal{P}_{\mathrm{cr}} \) and \( \tau_0 \omega_{\mathrm{p}} \) between approximately 1 and 2. However, these simplified analytical models cannot capture the full nonlinear structure of the objective landscape and therefore fail to identify the optimal LWFA operating regime revealed by the present work.

\bibliographystyle{unsrt}

\begin{thebibliography}{10}

\bibitem{tajima1979}
T.~Tajima and J.~M. Dawson.
\newblock Laser electron accelerator.
\newblock {\em Physical Review Letters}, 43(4):267--270, 1979.

\bibitem{esarey2009a}
E.~Esarey, C.~B. Schroeder, and W.~P. Leemans.
\newblock Physics of laser-driven plasma-based electron accelerators.
\newblock {\em Reviews of Modern Physics}, 81(3):1229--1285, August 2009.

\bibitem{aniculaesei2023}
C.~Aniculaesei, T.~Ha, S.~Yoffe, L.~Labun, S.~Milton, E.~McCary, M.~M. Spinks, H.~J. Quevedo, O.~Z. Labun, R.~Sain, A.~Hannasch, R.~Zgadzaj, I.~Pagano, J.~A. {Franco-Altamirano}, M.~L. Ringuette, E.~Gaul, S.~V. Luedtke, G.~Tiwari, B.~Ersfeld, E.~Brunetti, H.~Ruhl, T.~Ditmire, S.~Bruce, M.~E. Donovan, M.~C. Downer, D.~A. Jaroszynski, and B.~M. Hegelich.
\newblock The acceleration of a high-charge electron bunch to 10 {{GeV}} in a 10-cm nanoparticle-assisted wakefield accelerator.
\newblock {\em Matter and Radiation at Extremes}, 9(1):014001, November 2023.

\bibitem{picksley2024}
A.~Picksley, J.~Stackhouse, C.~Benedetti, K.~Nakamura, H.~E. Tsai, R.~Li, B.~Miao, J.~E. Shrock, E.~Rockafellow, H.~M. Milchberg, C.~B. Schroeder, J.~{van Tilborg}, E.~Esarey, C.~G.~R. Geddes, and A.~J. Gonsalves.
\newblock Matched {{Guiding}} and {{Controlled Injection}} in {{Dark-Current-Free}}, 10-{{GeV-Class}}, {{Channel-Guided Laser-Plasma Accelerators}}.
\newblock {\em Physical Review Letters}, 133(25):255001, December 2024.

\bibitem{corde2013a}
S.~Corde, K.~Ta~Phuoc, G.~Lambert, R.~Fitour, V.~Malka, A.~Rousse, A.~Beck, and E.~Lefebvre.
\newblock Femtosecond x rays from laser-plasma accelerators.
\newblock {\em Reviews of Modern Physics}, 85(1):1--48, January 2013.

\bibitem{albert2014}
F.~Albert, A.~G.~R. Thomas, S.~P.~D. Mangles, S.~Banerjee, S.~Corde, A.~Flacco, M.~Litos, D.~Neely, J.~Vieira, Z.~Najmudin, R.~Bingham, C.~Joshi, and T.~Katsouleas.
\newblock Laser wakefield accelerator based light sources: Potential applications and requirements.
\newblock {\em Plasma Physics and Controlled Fusion}, 56(8):084015, July 2014.

\bibitem{zhang2025}
F.~Zhang, L.~Deng, Y.~Ge, J.~Wen, B.~Cui, K.~Feng, H.~Wang, C.~Wu, Z.~Pan, H.~Liu, Z.~Deng, Z.~Zhang, L.~Chen, D.~Yan, L.~Shan, Z.~Yuan, C.~Tian, J.~Qian, J.~Zhu, Y.~Xu, Y.~Yu, X.~Zhang, L.~Yang, W.~Zhou, Y.~Gu, W.~Wang, Y.~Leng, Z.~Sun, and R.~Li.
\newblock Proof-of-principle demonstration of muon production with an ultrashort high-intensity laser.
\newblock {\em Nature Physics}, 21:1050--1056, May 2025.

\bibitem{ludwig2025}
J.~D. Ludwig, S.~C. Wilks, A.~J. Kemp, G.~J. Williams, N.~Lemos, E.~Rockafellow, B.~Miao, J.~E. Shrock, H.~M. Milchberg, J.-L. Vay, A.~Huebl, R.~Lehe, A.~Cimmino, R.~Versaci, S.~V. Bulanov, P.~Valenta, V.~Tang, and B.~A. Reagan.
\newblock Laser based 100 {{GeV}} electron acceleration scheme for muon production.
\newblock {\em Scientific Reports}, 15(1):25902, July 2025.

\bibitem{terzani2025a}
D.~Terzani, S.~Kisyov, S.~Greenberg, L.~Le~Pottier, M.~Mironova, A.~Picksley, J.~Stackhouse, H.-E. Tsai, R.~Li, E.~Rockafellow, B.~Miao, J.~E. Shrock, T.~Heim, M.~{Garcia-Sciveres}, C.~Benedetti, J.~Valentine, H.~M. Milchberg, K.~Nakamura, A.~J. Gonsalves, J.~{van Tilborg}, C.~B. Schroeder, E.~Esarey, and C.~G.~R. Geddes.
\newblock Measurement of directional muon beams generated at the berkeley lab laser accelerator.
\newblock {\em Physical Review Accelerators and Beams}, 28(10):103401, October 2025.

\bibitem{svendsen2021a}
K.~Svendsen, D.~Gu{\'e}not, J.~B. Svensson, K.~Petersson, A.~Persson, and O.~Lundh.
\newblock A focused very high energy electron beam for fractionated stereotactic radiotherapy.
\newblock {\em Scientific Reports}, 11(1):5844, March 2021.

\bibitem{horvath2023}
D.~Horv{\'a}th, G.~Grittani, M.~Precek, R.~Versaci, S.~V. Bulanov, and V.~Ol{\v s}ovcov{\'a}.
\newblock Time dynamics of the dose deposited by relativistic ultra-short electron beams.
\newblock {\em Physics in Medicine \& Biology}, 68(22):22NT01, November 2023.

\bibitem{guo2025}
Z.~Guo, S.~Liu, B.~Zhou, J.~Liu, H.~Wang, Y.~Pi, X.~Wang, Y.~Mo, B.~Guo, J.~Hua, Y.~Wan, and W.~Lu.
\newblock Preclinical tumor control with a laser-accelerated high-energy electron radiotherapy prototype.
\newblock {\em Nature Communications}, 16(1):1895, February 2025.

\bibitem{gonoskov2022}
A.~Gonoskov, T.~G. Blackburn, M.~Marklund, and S.~S. Bulanov.
\newblock Charged particle motion and radiation in strong electromagnetic fields.
\newblock {\em Reviews of Modern Physics}, 94(4):045001, October 2022.

\bibitem{russell2023}
B.~K. Russell, P.~T. Campbell, Q.~Qian, J.~A. Cardarelli, S.~S. Bulanov, S.~V. Bulanov, G.~M. Grittani, D.~Seipt, L.~Willingale, and A.~G.~R. Thomas.
\newblock Ultrafast relativistic electron probing of extreme magnetic fields.
\newblock {\em Physics of Plasmas}, 30(9):093105, September 2023.

\bibitem{dopp2023a}
A.~D{\"o}pp, C.~Eberle, S.~Howard, F.~Irshad, J.~Lin, and M.~Streeter.
\newblock Data-driven science and machine learning methods in laser--plasma physics.
\newblock {\em High Power Laser Science and Engineering}, 11:e55, January 2023.

\bibitem{roussel2024}
R.~Roussel, A.~L. Edelen, T.~Boltz, D.~Kennedy, Z.~Zhang, F.~Ji, X.~Huang, D.~Ratner, A.~S. Garcia, C.~Xu, J.~Kaiser, A.~Ferran~Pousa, A.~Eichler, J.~O. L{\"u}bsen, N.~M. Isenberg, Y.~Gao, N.~Kuklev, J.~Martinez, B.~Mustapha, V.~Kain, C.~Mayes, W.~Lin, S.~M. Liuzzo, J.~St.~John, M.~J.~V. Streeter, R.~Lehe, and W.~Neiswanger.
\newblock Bayesian optimization algorithms for accelerator physics.
\newblock {\em Physical Review Accelerators and Beams}, 27(8):084801, August 2024.

\bibitem{kostyukov2004}
I.~Kostyukov, A.~Pukhov, and S.~Kiselev.
\newblock Phenomenological theory of laser-plasma interaction in ``bubble'' regime.
\newblock {\em Physics of Plasmas}, 11(11):5256--5264, November 2004.

\bibitem{gordienko2005}
S.~Gordienko and A.~Pukhov.
\newblock Scalings for ultrarelativistic laser plasmas and quasimonoenergetic electrons.
\newblock {\em Physics of Plasmas}, 12(4):043109, April 2005.

\bibitem{lu2006}
W.~Lu, C.~Huang, M.~Zhou, W.~B. Mori, and T.~Katsouleas.
\newblock Nonlinear theory for relativistic plasma wakefields in the blowout regime.
\newblock {\em Physical Review Letters}, 96(16):165002, April 2006.

\bibitem{lu2007}
W.~Lu, M.~Tzoufras, C.~Joshi, F.~S. Tsung, W.~B. Mori, J.~Vieira, R.~A. Fonseca, and L.~O. Silva.
\newblock Generating multi-{{GeV}} electron bunches using single stage laser wakefield acceleration in a {{3D}} nonlinear regime.
\newblock {\em Physical Review Special Topics - Accelerators and Beams}, 10(6):61301, June 2007.

\bibitem{nedorezov2021}
V.~G. Nedorezov, S.~G. Rykovanov, and A.~B. Savel'ev.
\newblock Nuclear photonics: Results and prospects.
\newblock {\em Physics-Uspekhi}, 64(12):1214--1237, December 2021.

\bibitem{kolenaty2022}
D.~Kolenat{\'y}, P.~Hadjisolomou, R.~Versaci, T.~M. Jeong, P.~Valenta, V.~Ol{\v s}ovcov{\'a}, and S.~V. Bulanov.
\newblock Electron-positron pairs and radioactive nuclei production by irradiation of high-{{Z}} target with gamma-photon flash generated by an ultra-intense laser in the lambda{\textasciicircum}3 regime.
\newblock {\em Physical Review Research}, 4(2):023124, May 2022.

\bibitem{lifschitz2009a}
A.~F. Lifschitz, X.~Davoine, E.~Lefebvre, J.~Faure, C.~Rechatin, and V.~Malka.
\newblock Particle-in-{{Cell}} modelling of laser--plasma interaction using {{Fourier}} decomposition.
\newblock {\em Journal of Computational Physics}, 228(5):1803--1814, March 2009.

\bibitem{davidson2015}
A.~Davidson, A.~Tableman, W.~An, F.~S. Tsung, W.~Lu, J.~Vieira, R.~A. Fonseca, L.~O. Silva, and W.~B. Mori.
\newblock Implementation of a hybrid particle code with a {{PIC}} description in r--z and a gridless description in {$\phi$} into {{OSIRIS}}.
\newblock {\em Journal of Computational Physics}, 281:1063--1077, 2015.

\bibitem{vay2007}
J.-L. Vay.
\newblock Noninvariance of {{Space-}} and {{Time-Scale Ranges}} under a {{Lorentz Transformation}} and the {{Implications}} for the {{Study}} of {{Relativistic Interactions}}.
\newblock {\em Physical Review Letters}, 98(13):130405, March 2007.

\bibitem{fonseca2008}
R.~A. Fonseca, S.~F. Martins, L.~O. Silva, J.~W. Tonge, F.~S. Tsung, and W.~B. Mori.
\newblock One-to-one direct modeling of experiments and astrophysical scenarios: Pushing the envelope on kinetic plasma simulations.
\newblock {\em Plasma Physics and Controlled Fusion}, 50(12):124034, November 2008.

\bibitem{yu2016}
P.~Yu, X.~Xu, A.~Davidson, A.~Tableman, T.~Dalichaouch, F.~Li, M.~D. Meyers, W.~An, F.~S. Tsung, V.~K. Decyk, F.~Fiuza, J.~Vieira, R.~A. Fonseca, W.~Lu, L.~O. Silva, and B.~Mori.
\newblock Enabling {{Lorentz}} boosted frame particle-in-cell simulations of laser wakefield acceleration in quasi-{{3D}} geometry.
\newblock {\em Journal of Computational Physics}, 316:747--759, July 2016.

\bibitem{shalloo2020}
R.~J. Shalloo, S.~J.~D. Dann, J.-N. Gruse, C.~I.~D. Underwood, A.~F. Antoine, C.~Arran, M.~Backhouse, C.~D. Baird, M.~D. Balcazar, N.~Bourgeois, J.~A. Cardarelli, P.~Hatfield, J.~Kang, K.~Krushelnick, S.~P.~D. Mangles, C.~D. Murphy, N.~Lu, J.~Osterhoff, K.~P{\~o}der, P.~P. Rajeev, C.~P. Ridgers, S.~Rozario, M.~P. Selwood, A.~J. Shahani, D.~R. Symes, A.~G.~R. Thomas, C.~Thornton, Z.~Najmudin, and M.~J.~V. Streeter.
\newblock Automation and control of laser wakefield accelerators using {{Bayesian}} optimization.
\newblock {\em Nature Communications}, 11(1):6355, December 2020.

\bibitem{jalas2021}
S.~Jalas, M.~Kirchen, P.~Messner, P.~Winkler, L.~H{\"u}bner, J.~Dirkwinkel, M.~Schnepp, R.~Lehe, and A.~R. Maier.
\newblock Bayesian {{Optimization}} of a {{Laser-Plasma Accelerator}}.
\newblock {\em Physical Review Letters}, 126(10):104801, March 2021.

\bibitem{kirchen2021}
M.~Kirchen, S.~Jalas, P.~Messner, P.~Winkler, T.~Eichner, L.~H{\"u}bner, T.~H{\"u}lsenbusch, L.~Jeppe, T.~Parikh, M.~Schnepp, and A.~R. Maier.
\newblock Optimal {{Beam Loading}} in a {{Laser-Plasma Accelerator}}.
\newblock {\em Physical Review Letters}, 126(17):174801, April 2021.

\bibitem{irshad2024}
F.~Irshad, C.~Eberle, F.~M. Foerster, K.~v.~Grafenstein, F.~Haberstroh, E.~Travac, N.~Weisse, S.~Karsch, and A.~D{\"o}pp.
\newblock Pareto {{Optimization}} and {{Tuning}} of a {{Laser Wakefield Accelerator}}.
\newblock {\em Physical Review Letters}, 133(8):085001, August 2024.

\bibitem{valenta2025b}
P.~Valenta, T.~{\relax Zh}. Esirkepov, J.~D. Ludwig, S.~C. Wilks, and S.~V. Bulanov.
\newblock Bayesian optimization of electron energy from laser wakefield accelerators.
\newblock {\em Physical Review Accelerators and Beams}, 28(9):094601, September 2025.

\bibitem{martins2010}
S.~F. Martins, R.~A. Fonseca, W.~Lu, W.~B. Mori, and L.~O. Silva.
\newblock Exploring laser-wakefield-accelerator regimes for near-term lasers using particle-in-cell simulation in {{Lorentz-boosted}} frames.
\newblock {\em Nature Physics}, 6(4):311--316, April 2010.

\bibitem{yu2014}
P.~Yu, X.~Xu, V.~K. Decyk, W.~An, J.~Vieira, F.~S. Tsung, R.~A. Fonseca, W.~Lu, L.~O. Silva, and W.~B. Mori.
\newblock Modeling of laser wakefield acceleration in {{Lorentz}} boosted frame using {{EM-PIC}} code with spectral solver.
\newblock {\em Journal of Computational Physics}, 266:124--138, June 2014.

\bibitem{massimo2025}
F.~Massimo, C.~Benedetti, D.~Terzani, A.~Beck, and B.~Cros.
\newblock Modeling laser-wakefield accelerators using the time-averaged ponderomotive approximation in a {{Lorentz}} boosted frame.
\newblock {\em Plasma Physics and Controlled Fusion}, 67(6):065032, June 2025.

\bibitem{zeng2024}
M.~Zeng.
\newblock Simulation observation of high effectiveness laser plasma wakefield accelerator using plasma telescope configuration.
\newblock {\em Physics of Plasmas}, 31(8):080702, August 2024.

\bibitem{ehrlich1996}
Y.~Ehrlich, C.~Cohen, A.~Zigler, J.~Krall, P.~Sprangle, and E.~Esarey.
\newblock Guiding of {{High Intensity Laser Pulses}} in {{Straight}} and {{Curved Plasma Channel Experiments}}.
\newblock {\em Physical Review Letters}, 77(20):4186--4189, November 1996.

\bibitem{butler2002}
A.~Butler, D.~J. Spence, and S.~M. Hooker.
\newblock Guiding of high-intensity laser pulses with a hydrogen-filled capillary discharge waveguide.
\newblock {\em Physical Review Letters}, 89(18):185003, October 2002.

\bibitem{durfee1993}
C.~G. Durfee and H.~M. Milchberg.
\newblock Light pipe for high intensity laser pulses.
\newblock {\em Physical Review Letters}, 71(15):2409--2412, October 1993.

\bibitem{miao2020}
B.~Miao, L.~Feder, J.~E. Shrock, A.~Goffin, and H.~M. Milchberg.
\newblock Optical {{Guiding}} in {{Meter-Scale Plasma Waveguides}}.
\newblock {\em Physical Review Letters}, 125(7):074801, August 2020.

\bibitem{sun1987}
G.-Z. Sun, E.~Ott, Y.~C. Lee, and P.~Guzdar.
\newblock Self-focusing of short intense pulses in plasmas.
\newblock {\em Physics of Fluids}, 30(2):526, 1987.

\bibitem{naumova2002}
N.~M. Naumova, S.~V. Bulanov, K.~Nishihara, T.~{\relax Zh}. Esirkepov, and F.~Pegoraro.
\newblock Polarization effects and anisotropy in three-dimensional relativistic self-focusing.
\newblock {\em Physical Review E}, 65(4):045402, April 2002.

\bibitem{valenta2021a}
P.~Valenta, G.~M. Grittani, C.~M. Lazzarini, O.~Klimo, and S.~V. Bulanov.
\newblock On the electromagnetic-electron rings originating from the interaction of high-power short-pulse laser and underdense plasma.
\newblock {\em Physics of Plasmas}, 28(12):122104, December 2021.

\bibitem{bulanov1996}
S.~V. Bulanov, M.~Lontano, T.~{\relax Zh}. Esirkepov, F.~Pegoraro, and A.~M. Pukhov.
\newblock Electron vortices produced by ultraintense laser pulses.
\newblock {\em Physical Review Letters}, 76(19):3562--3565, May 1996.

\bibitem{bulanov1999}
S.~V. Bulanov, T.~{\relax Zh}. Esirkepov, N.~M. Naumova, F.~Pegoraro, and V.~A. Vshivkov.
\newblock Solitonlike electromagnetic waves behind a superintense laser pulse in a plasma.
\newblock {\em Physical Review Letters}, 82(17):3440--3443, April 1999.

\bibitem{esirkepov2002}
T.~{\relax Zh}. Esirkepov, K.~Nishihara, S.~V. Bulanov, and F.~Pegoraro.
\newblock Three-dimensional relativistic electromagnetic subcycle solitons.
\newblock {\em Physical Review Letters}, 89(27):275002, December 2002.

\bibitem{pukhov2002}
A.~Pukhov and J.~{Meyer-ter-Vehn}.
\newblock Laser wake field acceleration: {{The}} highly non-linear broken-wave regime.
\newblock {\em Applied Physics B}, 74(4-5):355--361, April 2002.

\bibitem{beaurepaire2015}
B.~Beaurepaire, A.~Vernier, M.~Bocoum, F.~B{\"o}hle, A.~Jullien, J-P. Rousseau, T.~Lefrou, D.~Douillet, G.~Iaquaniello, R.~{Lopez-Martens}, A.~Lifschitz, and J.~Faure.
\newblock Effect of the {{Laser Wave Front}} in a {{Laser-Plasma Accelerator}}.
\newblock {\em Physical Review X}, 5(3):031012, July 2015.

\bibitem{oumbarekespinos2023}
D.~Oumbarek~Espinos, A.~Rondepierre, A.~Zhidkov, N.~Pathak, Z.~Jin, K.~Huang, N.~Nakanii, I.~Daito, M.~Kando, and T.~Hosokai.
\newblock Notable improvements on {{LWFA}} through precise laser wavefront tuning.
\newblock {\em Scientific Reports}, 13(1):18466, October 2023.

\bibitem{kalmykov2012}
S.~Y. Kalmykov, A.~Beck, X.~Davoine, E.~Lefebvre, and B.~A. Shadwick.
\newblock Laser plasma acceleration with a negatively chirped pulse: All-optical control over dark current in the blowout regime.
\newblock {\em New Journal of Physics}, 14(3):033025, March 2012.

\bibitem{kim2017a}
H.~T. Kim, V.~B. Pathak, K.~Hong~Pae, A.~Lifschitz, F.~Sylla, J.~H. Shin, C.~Hojbota, S.~K. Lee, J.~H. Sung, H.~W. Lee, E.~Guillaume, C.~Thaury, .~Nakajima, J.~Vieira, L.~O. Silva, V.~Malka, and C.~H. Nam.
\newblock Stable multi-{{GeV}} electron accelerator driven by waveform-controlled {{PW}} laser pulses.
\newblock {\em Scientific Reports}, 7(1):10203, August 2017.

\bibitem{bulanov1993}
S.~V. Bulanov, V.~I. Kirsanov, F.~Pegoraro, and A.~S. Sakharov.
\newblock Charged particle and photon acceleration by wake field plasma waves in nonuniform plasmas.
\newblock {\em Laser Physics}, 3:1078, 1993.

\bibitem{sprangle2001}
P.~Sprangle, B.~Hafizi, J.~R. Pe{\~n}ano, R.~F. Hubbard, A.~Ting, C.~I. Moore, D.~F. Gordon, A.~Zigler, D.~Kaganovich, and T.~M. Antonsen.
\newblock Wakefield generation and {{GeV}} acceleration in tapered plasma channels.
\newblock {\em Physical Review E}, 63(5):056405, April 2001.

\bibitem{perevalov2020}
S.~E. Perevalov, K.~F. Burdonov, A.~V. Kotov, D.~S. Romanovskiy, A.~A. Soloviev, M.~V. Starodubtsev, A.~A. Golovanov, V.~N. Ginzburg, A.~A. Kochetkov, A.~P. Korobeinikova, A.~A. Kuz'min, I.~A. Shaikin, A.~A. Shaykin, I.~V. Yakovlev, E.~A. Khazanov, and I.~Yu Kostyukov.
\newblock Experimental study of strongly mismatched regime of laser-driven wakefield acceleration.
\newblock {\em Plasma Physics and Controlled Fusion}, 62(9):094004, August 2020.

\bibitem{poder2024}
K.~P{\~o}der, J.~C. Wood, N.~C. Lopes, J.~M. Cole, S.~Alatabi, M.~P. Backhouse, P.~S. Foster, A.~J. Hughes, C.~Kamperidis, O.~Kononenko, S.~P.~D. Mangles, C.~A.~J. Palmer, D.~Rusby, A.~Sahai, G.~Sarri, D.~R. Symes, J.~R. Warwick, and Z.~Najmudin.
\newblock Multi-{{GeV Electron Acceleration}} in {{Wakefields Strongly Driven}} by {{Oversized Laser Spots}}.
\newblock {\em Physical Review Letters}, 132(19):195001, May 2024.

\bibitem{stark2021}
H.~Stark, J.~Buldt, M.~M{\"u}ller, A.~Klenke, and J.~Limpert.
\newblock 1 {{kW}}, 10 {{mJ}}, 120 fs coherently combined fiber {{CPA}} laser system.
\newblock {\em Optics Letters}, 46(5):969--972, March 2021.

\bibitem{nerush2009}
E.~N. Nerush and I.~{\relax Yu}. Kostyukov.
\newblock Carrier-envelope phase effects in plasma-based electron acceleration with few-cycle laser pulses.
\newblock {\em Physical Review Letters}, 103(3):035001, July 2009.

\bibitem{valenta2020}
P.~Valenta, T.~Zh Esirkepov, J.~K. Koga, A.~Ne{\v c}as, G.~M. Grittani, C.~M. Lazzarini, O.~Klimo, G.~Korn, and S.~V. Bulanov.
\newblock Polarity reversal of wakefields driven by ultrashort pulse laser.
\newblock {\em Physical Review E}, 102(5):53216, November 2020.

\bibitem{salehi2021a}
F.~Salehi, M.~Le, L.~Railing, M.~Kolesik, and H.~M. Milchberg.
\newblock Laser-{{Accelerated}}, {{Low-Divergence}} 15-{{MeV Quasimonoenergetic Electron Bunches}} at 1 {{kHz}}.
\newblock {\em Physical Review X}, 11(2):021055, June 2021.

\bibitem{huijts2022}
J.~Huijts, L.~Rovige, I.~A. Andriyash, A.~Vernier, M.~Ouill{\'e}, J.~Kaur, Z.~Cheng, R.~{Lopez-Martens}, and J.~Faure.
\newblock Waveform {{Control}} of {{Relativistic Electron Dynamics}} in {{Laser-Plasma Acceleration}}.
\newblock {\em Physical Review X}, 12(1):011036, February 2022.

\bibitem{lazzarini2024}
C.~M. Lazzarini, G.~M. Grittani, P.~Valenta, I.~Zymak, R.~Antipenkov, U.~Chaulagain, L.~V.~N. Goncalves, A.~Grenfell, M.~Lama{\v c}, S.~Lorenz, M.~Nevrkla, A.~{\v S}pa{\v c}ek, V.~{\v S}obr, W.~Szuba, P.~Bakule, G.~Korn, and S.~V. Bulanov.
\newblock Ultrarelativistic electron beams accelerated by terawatt scalable {{kHz}} laser.
\newblock {\em Physics of Plasmas}, 31(3):030703, March 2024.

\bibitem{wilks1987}
S.~Wilks, T.~Katsouleas, J.~M. Dawson, P.~Chen, and J.~J. Su.
\newblock Beam {{Loading}} in {{Plasma Waves}}.
\newblock {\em IEEE Transactions on Plasma Science}, 15(2):210--217, April 1987.

\bibitem{katsouleas1987}
T.~Katsouleas, S.~Wilks, P.~Chen, J.~M. Dawson, and J.~J. Su.
\newblock Beam loading in plasma accelerators.
\newblock {\em Particle Accelerators}, 22(1):81--99, 1987.

\bibitem{cho2018}
M.~H. Cho, V.~B. Pathak, H.~T. Kim, and C.~H. Nam.
\newblock Controlled electron injection facilitated by nanoparticles for laser wakefield acceleration.
\newblock {\em Scientific Reports}, 8(1):16924, November 2018.

\bibitem{spadova2025}
A.~{\v S}p{\'a}dov{\'a}, P.~Valenta, S.~Lorenz, M.~Nevrkla, J.~Nejdl, G.~M. Grittani, and S.~V. Bulanov.
\newblock Toward controlling electron beam charge with nanoparticle-assisted laser wakefield accelerators.
\newblock {\em Physics of Plasmas}, 32(12):123104, December 2025.

\bibitem{fonseca2002}
R.~A. Fonseca, L.~O. Silva, F.~S. Tsung, V.~K. Decyk, W.~Lu, C.~Ren, W.~B. Mori, S.~Deng, S.~Lee, T.~Katsouleas, and J.~C. Adam.
\newblock {{OSIRIS}}: {{A}} three-dimensional, fully relativistic particle in cell code for modeling plasma based accelerators.
\newblock Computational {{Science}} --- {{ICCS}} 2002, pages 342--351. Springer Berlin Heidelberg, 2002.

\bibitem{boris1971}
J.~P. Boris.
\newblock Relativistic plasma simulation - {{Optimization}} of a hybrid code.
\newblock In J.~P. Boris and R.~A. Shanny, editors, {\em Proceedings of the Fourth Conference on Numerical Simulation of Plasmas}, pages 3--67. Naval Research Laboratory, 1971.

\bibitem{li2017}
F.~Li, P.~Yu, X.~Xu, F.~Fiuza, V.~K. Decyk, T.~Dalichaouch, A.~Davidson, A.~Tableman, W.~An, F.~S. Tsung, R.~A. Fonseca, W.~Lu, and W.~B. Mori.
\newblock Controlling the numerical {{Cerenkov}} instability in {{PIC}} simulations using a customized finite difference {{Maxwell}} solver and a local {{FFT}} based current correction.
\newblock {\em Computer Physics Communications}, 214:6--17, 2017.

\bibitem{hudson2022}
S.~Hudson, J.~Larson, J.-L. Navarro, and S.~M. Wild.
\newblock {{libEnsemble}}: {{A Library}} to {{Coordinate}} the {{Concurrent Evaluation}} of {{Dynamic Ensembles}} of {{Calculations}}.
\newblock {\em IEEE Transactions on Parallel and Distributed Systems}, 33(4):977--988, April 2022.

\bibitem{ferranpousa2023}
A.~Ferran~Pousa, S.~Jalas, M.~Kirchen, A.~{Martinez de la Ossa}, M.~Th{\'e}venet, S.~Hudson, J.~Larson, A.~Huebl, J.-L. Vay, and R.~Lehe.
\newblock Bayesian optimization of laser-plasma accelerators assisted by reduced physical models.
\newblock {\em Physical Review Accelerators and Beams}, 26(8):084601, August 2023.

\bibitem{rasmussen2005}
C.~E. Rasmussen and C.~K.~I. Williams.
\newblock {\em Gaussian {{Processes}} for {{Machine Learning}}}.
\newblock The MIT Press, November 2005.

\bibitem{powell1977}
M.~J.~D. Powell.
\newblock Restart procedures for the conjugate gradient method.
\newblock {\em Mathematical Programming}, 12(1):241--254, December 1977.

\bibitem{auer2002}
P.~Auer, N.~{Cesa-Bianchi}, and P.~Fischer.
\newblock Finite-time {{Analysis}} of the {{Multiarmed Bandit Problem}}.
\newblock {\em Machine Learning}, 47(2):235--256, May 2002.

\bibitem{wilson2017}
J.~T. Wilson, R.~Moriconi, F.~Hutter, and M.~P. Deisenroth.
\newblock The reparameterization trick for acquisition functions.
\newblock {\em arXiv:1712.00424}, December 2017.

\bibitem{gong2023}
Z.~Gong, M.~J. Quin, S.~Bohlen, C.~H. Keitel, K.~P{\~o}der, and M.~Tamburini.
\newblock Spin-polarized electron beam generation in the colliding-pulse injection scheme.
\newblock {\em Matter and Radiation at Extremes}, 8(6):064005, October 2023.

\bibitem{clarke1994}
L.~Clarke, I.~Glendinning, and R.~Hempel.
\newblock The {{MPI Message Passing Interface Standard}}.
\newblock In K.~M. Decker and R.~M. Rehmann, editors, {\em Programming {{Environments}} for {{Massively Parallel Distributed Systems}}}, pages 213--218, Basel, 1994. Birkh{\"a}user.

\bibitem{sobol1967}
I.~M. Sobol'.
\newblock On the distribution of points in a cube and the approximate evaluation of integrals.
\newblock {\em USSR Computational Mathematics and Mathematical Physics}, 7(4):86--112, January 1967.

\bibitem{leemans1996}
W.~P. Leemans, C.~W. Siders, E.~Esarey, N.~E. Andreev, G.~Shvets, and W.~B. Mori.
\newblock Plasma guiding and wakefield generation for second-generation experiments.
\newblock {\em IEEE Transactions on Plasma Science}, 24(2):331--342, April 1996.

\bibitem{valenta2026_zenodo}
P.~Valenta.
\newblock Optimized matching conditions for self-guided laser wakefield accelerators [{{Data}} set].
\newblock {\em Zenodo}, https://doi.org/10.5281/zenodo.18610970, February 2026.

\end{thebibliography}

\end{document}